\RequirePackage{fix-cm}
\documentclass[twocolumn,epjc3]{svjour3}  % EPJC
\smartqed  % flush right qed marks, e.g. at end of proof

\usepackage{graphicx}  % needed for figures
\usepackage{dcolumn}   % needed for some tables
\usepackage{bm}        % for math
\usepackage{amssymb}   % for math
\usepackage{sidecap}   % sidecaptions
\usepackage{subfigure}
\usepackage{amsmath}
\usepackage{pdfpages} % for authorlist
\usepackage{multirow}
\usepackage{xspace}
\usepackage{booktabs}
\usepackage[normalem]{ulem}
\usepackage[numbers,sort&compress]{natbib}
\usepackage{epstopdf} %allows to use pdftex

\usepackage{hyperref}

\usepackage{atlasphysics} 
\usepackage{subfigure}

\usepackage{preprintcover}  % load package
\PreprintCoverPaperTitle{Measurement of $t\bar{t}$~production with a veto on additional central jet activity in $pp$~collisions at $\sqrt{s}=7$~TeV using the ATLAS detector}  % paper title
\PreprintIdNumber{CERN-PH-EP-2012-062}  % CERN preprint number
\PreprintCoverAbstract{A measurement of the jet activity in $t\bar{t}$ events produced in proton-proton collisions at a centre-of-mass energy of $7$~TeV is presented, using $2.05$~\ifb\ of integrated luminosity collected by the ATLAS detector at the Large Hadron Collider. 
The $t\bar{t}$ events are selected in the dilepton decay channel with two identified $b$-jets from the top quark decays. Events are vetoed if they contain an additional jet with transverse momentum above a threshold in a central rapidity interval. The fraction of events surviving the jet veto is presented as a function of this threshold for four different central rapidity interval definitions. 
An alternate measurement is also performed, in which events are vetoed if the scalar transverse momentum sum of the additional jets in each rapidity interval is above a threshold. 
In both measurements, the data are corrected for detector effects and compared to the theoretical models implemented in \mcatnlo{}, \powheg{}, \alpgen{} and \sherpa{}. 
The experimental uncertainties are often smaller than the spread of theoretical predictions, allowing deviations between data and theory to be observed in some regions of phase space. 
}  % paper abstract
\PreprintJournalName{Eur. Phys. J. C}  % journal name

\def\Qz{\ensuremath{Q_0}}
\def\Qsum{\ensuremath{Q_{\rm sum}}}
\def\pythia{{\sc Pythia}}
\def\sherpa{{\sc Sherpa}}
\def\alpgen{{\sc Alpgen}}
\def\powheg{{\sc Powheg}}
\def\herwig{{\sc Herwig}}
\def\jimmy{{\sc Jimmy}}
\def\acermc{{\sc AcerMC}}
\def\mcatnlo{{\sc MC@NLO}}

\journalname{Eur.\ Phys.\ J.\ C}

\begin{document}

\title{Measurement of $t\bar{t}$~production with a veto on additional central jet activity in $pp$~collisions at $\sqrt{s}=7$~TeV using the ATLAS detector}

\author{The ATLAS Collaboration}
\institute{CERN, 1211 Geneva 23, Switzerland}
\date{April 19, 2012}

\maketitle % for EPJC / RevTex

\begin{abstract}
A measurement of the jet activity in $t\bar{t}$ events produced in proton-proton collisions at a centre-of-mass energy of $7$~TeV is presented, using $2.05$~\ifb\ of integrated luminosity collected by the ATLAS detector at the Large Hadron Collider. 
The $t\bar{t}$ events are selected in the dilepton decay channel with two identified $b$-jets from the top quark decays. Events are vetoed if they contain an additional jet with transverse momentum above a threshold in a central rapidity interval. The fraction of events surviving the jet veto is presented as a function of this threshold for four different central rapidity interval definitions. 
An alternate measurement is also performed, in which events are vetoed if the scalar transverse momentum sum of the additional jets in each rapidity interval is above a threshold. 
In both measurements, the data are corrected for detector effects and compared to the theoretical models implemented in \mcatnlo{}, \powheg{}, \alpgen{} and \sherpa{}. 
The experimental uncertainties are often smaller than the spread of theoretical predictions, allowing deviations between data and theory to be observed in some regions of phase space. 
%PACS - top quark, experimental tests of QCD, inclusive high energy hadrons with leptons
\end{abstract}

\section{Introduction}
\label{sec-intro}

Measurements of the top quark provide an important test of the Standard Model (SM) and any observed deviation from the SM predictions could indicate the presence of new physics. However, many top quark measurements have large uncertainties that arise from the theoretical description of quark and gluon radiation in the standard Monte Carlo (MC) event generators. Recent measurements that are affected by such modelling uncertainties include the $t\bar{t}$ production cross-section \cite{Aad:2011yb,Aad:2012qf,inreview1,Chatrchyan:2011nb}, the spin correlations in $\ttbar$~events~\cite{Collaboration:2012sm}, the charge asymmetry \cite{inreview3,Chatrchyan:2011hk} and the top quark mass \cite{Chatrchyan:2011nb}.
In addition, a significant disagreement between data and the prediction from \mcatnlo~\cite{Frixione:2002ik,Frixione:2003ei} was observed by the D0 Collaboration in the  transverse momentum distribution of the $\ttbar$~system \cite{Abazov:2011rq}. This disagreement obscures the interpretation of the observed forward-backward asymmetry in terms of a deviation from SM predictions. 
Measurements sensitive to the theoretical description of quark and gluon radiation in events containing a $\ttbar$ final state are therefore needed in order to constrain the modelling and reduce the impact on future experimental measurements.

In this article, a jet veto is used to quantify the jet activity that arises from quark and gluon radiation produced in association with the $t\bar{t}$ system.  The events are selected in the dilepton decay channel so that the additional jets can be easily distinguished from the $t\bar{t}$ decay products (two leptons and two jets originating from $b$-quarks). The variable of interest is the `gap fraction', defined as
\begin{equation}
f(\Qz) = \frac{n(\Qz)}{N},
\label{e:gf}
\end{equation}
where $N$~is the number of selected $\ttbar$~events and $n(\Qz)$~is the subset of these events that do not contain an additional jet with transverse momentum, $p_{\rm T}$, above a threshold, $\Qz$, in a central rapidity\footnote{
ATLAS uses a right-handed coordinate system with the $z$-axis along the beam line. 
Cylindrical coordinates ($r$, $\phi$) are used in the
transverse plane, $\phi$~being the azimuthal angle. Pseudorapidity is defined in terms of the polar
angle $\theta$~as $\eta = - {\mathrm{ln}}\left[\tan(\theta/2)\right]$.
Rapidity is defined as $y = 0.5\,\ln\left[(E + p_z)/(E - p_z)\right]$~where
$E$\ denotes the energy and $p_z$\ is the component of the momentum along the beam direction.
Transverse momentum and energy are defined as $\pt = p \sin \theta$~and $\et = E \sin \theta$, respectively.
} interval. The minimum jet $\pt$ used in the measurement is $25$~GeV. The measurement is corrected for detector effects and presented in a fiducial region. The gap fraction can then be written as
\begin{equation}
f(\Qz) = \frac{\sigma(Q_0)}{\sigma},
\label{e:gfsig}
\end{equation}
where $\sigma$~is the fiducial cross section for inclusive $\ttbar$ production and $\sigma(\Qz )$ is the fiducial cross section for $\ttbar$ events produced in the absence of an additional jet with $\pt > \Qz$~in the rapidity interval. The gap fraction is measured for multiple values of $\Qz$ and for four jet rapidity intervals: $|y|<0.8$, $0.8 \leq |y|<1.5$, $1.5\leq |y|<2.1$ and $|y|<2.1$. 

The veto criterion can be extended to probe jet activity beyond the leading additional jet. An alternate definition of the gap fraction is used in this case,
\begin{equation}
f(\Qsum) = \frac{n(\Qsum)}{N} \equiv \frac{\sigma(Q_{\rm sum})}{\sigma},
\label{e:gf2}
\end{equation}
where $n(\Qsum)$ is the number of $\ttbar$~events, and $\sigma(\Qsum)$ is the cross section, in which the scalar  transverse momentum sum of the additional jets in the rapidity interval is less than $\Qsum$. The gap fraction defined using $\Qz$ is mainly sensitive to the leading-$\pt$~emission accompanying the $\ttbar$~system, whereas the gap fraction defined using $\Qsum$ is sensitive to all hard
emissions accompanying the $\ttbar$~system.

Many of the experimental systematic uncertainties cancel in the ratio, as observed in the ATLAS measurement of the gap fraction in dijet events \cite{Aad:2011jz}. 
The data are therefore expected to constrain the modelling of quark and gluon radiation in $\ttbar$~events and provide useful information about the general theoretical description of jet vetoes, which have been proposed as a tool to enhance new physics signals~\cite{Barger:1990py,Barger:1991ar,Barger:1994zq}, and to study the properties of new fundamental particles~\cite{Cox:2010ug,Sung:2009iq,Ask:2011zs}.

\section{ATLAS Detector}
\label{sec-atlas}

The ATLAS detector~\cite{Aad:2008zzm} surrounds one of the proton-proton interaction points at the Large Hadron Collider. The inner tracking detector is composed of silicon pixel detectors,
silicon microstrip detectors and a transition radiation tracking detector. The inner detector is surrounded by a superconducting solenoid that provides a $2$~T magnetic field. This allows the momentum of charged particles that pass through the inner detector to be determined for $|\eta|<2.5$. Outside the solenoid are liquid-argon electromagnetic sampling calorimeters ($|\eta|<3.2$). Hadronic energy measurements are
provided by a scintillator tile calorimeter in the central region ($|\eta| < 1.7$)
 and by liquid-argon calorimetry up to $|\eta|<4.9$.
The muon spectrometer system surrounds the calorimeter system and incorporates a toroidal magnet system, with a field of approximately $0.5$~and $1$~T in the barrel
and endcap regions respectively. The muon spectrometer provides precision measurements of the momentum of muons up to $|\eta|<2.7$, while the corresponding trigger chambers
are limited to $|\eta|<2.4$. 

The data are collected using a three-level trigger system. The first level is implemented in hardware and reduces the data rate to less than $75$~kHz. The following two software trigger levels reduce the rate to several hundred Hz. The data passing the trigger selections are recorded for use in subsequent analyses.

The measurements presented in this paper use data from proton-proton collisions at a centre-of-mass energy $\sqrt{s}=7$~TeV, and rely on triggers designed to select events that contain high transverse momentum electrons or muons. The integrated luminosity of the data sample is $2.05\pm0.08$~fb$^{-1}$  \cite{Aad:2011dr,ATLAS-CONF-2011-116}.

\section{Theoretical Predictions}
\label{sec-theory}

The theoretical predictions for $\ttbar$~production are produced using the \mcatnlo{} \cite{Frixione:2002ik,Frixione:2003ei}, \powheg{} \cite{Nason:2004rx,Frixione:2007vw}, \alpgen{} \cite{ref:noteALPGEN}, \sherpa{} \cite{Gleisberg:2008ta} and \acermc{} \cite{KER-0401,Stelzer:1994ta} event generators. 

\mcatnlo~provides a calculation of $t\bar{t}$ production at next-to-leading order (NLO) accuracy and is interfaced to \herwig~\cite{COR-0001} and \jimmy~\cite{Butterworth:1996zw} for parton showering, hadronisation and underlying event from multiple partonic interactions. The parton distribution functions (PDF) chosen to generate the \mcatnlo~ events are CTEQ6.6 \cite{Pumplin:2002vw} and the underlying event tune for \herwig/\jimmy{} is chosen to be AUET1 \cite{auet}. 
\powheg{} also produces the $t\bar{t}$ final state to NLO accuracy using the CTEQ6.6 PDF. The parton showering, hadronisation and underlying event are added by interfacing to either \pythia{} \cite{Pythia}, with underlying event tune AMBT1 \cite{ATLAS:1266235}, or to \herwig/\jimmy{}, with underlying event tune AUET1. 

\alpgen{} provides leading order (LO) matrix elements for $t\bar{t}$ production with up to three additional partons in the final state. The \alpgen{} events are generated using the CTEQ6L1 PDF \cite{Pumplin:2002vw} and interfaced to \herwig/\jimmy{}  for parton showering, hadronisation and underlying event (tune AUET1). The MLM matching procedure \cite{Mangano:2006rw} is used to remove double counting between partons produced by the matrix element and parton shower.
\sherpa{} is also used to generate $t\bar{t}$ events with up to three additional partons in the final state. This provides an independent LO matrix-element calculation with a different matching scheme (CKKW \cite{Catani:2001cc}) between the matrix element and the parton shower. The events are generated with the default underlying event tune and the CTEQ6L1 PDF.

\acermc{} consists of a LO matrix element for $t\bar{t}$ production and is interfaced to \pythia{} to provide the hadronic final state, using the MRST2007LO$^{*}$ PDF \cite{Sherstnev:2007nd} and underlying event tune AMBT1. Three samples are produced with nominal, increased and decreased initial state radiation (ISR)\footnote{The default ISR parameters in AMBT1 are
PARP(67)=4.0 and PARP(64)=1.0. To decrease ISR, the parameters are set to
 0.5 and 4.0, respectively. To increase ISR, they are set to 6.0 and 0.25, respectively.}. These samples have been previously used to assess ISR-based modelling uncertainties in ATLAS top quark measurements \cite{Aad:2011yb,Aad:2012qf,inreview1,Collaboration:2012sm,inreview3}.

\section{Simulation Samples}
\label{sec-mcsamples}

In order to simulate the events observed in the detector, several MC samples are passed through
the GEANT4~\cite{Agostinelli:2002hh} simulation of the ATLAS detector~\cite{:2010wqa} and
are processed with the same reconstruction chain as used for the data.
The \mcatnlo~and \powheg~samples described in Section~\ref{sec-theory} are used
to simulate the $\ttbar$~events. The background contribution from 
single top, $Z$+jets and diboson production is estimated using \mcatnlo{} \cite{Frixione:2008yi}, \alpgen\ and
\herwig, respectively. The hadronic final state for each of these backgrounds is generated using \herwig{}/\jimmy{} with underlying event tune AUET1. The MC samples are overlaid with additional minimum bias events generated with
\pythia{} to simulate the effect of additional proton-proton interactions. The simulated events are 
re-weighted such that the average number of interactions  per proton-proton bunch crossing, $\langle \mu \rangle$, is the same in data and MC simulation. This average varies between data-taking periods and is typically in the range $4 < \langle \mu \rangle < 8$. 

Corrections are applied to the simulation to reflect the observed performance in the data.
The electron reconstruction efficiency, energy scale and energy resolution
 are corrected to match the observed distributions in $W\rightarrow e\nu$ and $Z\rightarrow ee$\ events~\cite{Aad:2011mk}. The muon reconstruction efficiency, momentum scale and momentum resolution
 are corrected to match the observation in $Z\rightarrow \mu\mu$~events.
 The jet energy resolution is found to be larger in the data than predicted by the simulation and additional smearing is  applied to the simulated jets to ensure the resolution matches that in the data.
 The efficiency and rejection rate of the algorithm used to identify jets that have originated from $b$-quarks
 is measured in the data and the simulation is corrected on a per-jet basis
 to match the observed performance.
 All these corrections have associated systematic uncertainties and the effect of these on the
 measurement of the gap fraction is discussed in Section~\ref{sec-syst}.

\section{Event Selection}
\label{sec-eventsel}

The selection of $\ttbar$~events closely follows the selection used in the recent measurement of the $\ttbar$~production cross section~\cite{inreview1}. Electrons are required to have transverse energy $E_{\rm T}>25$ GeV and $|\eta|<2.47$, whereas muons
are required to have $\pt > 20$~GeV and $|\eta|<2.5$.
Electrons in the transition region between the barrel and endcap calorimeters ($1.37 < |\eta| < 1.52$) are excluded.

Jets are reconstructed using the anti-$k_{\rm t}$ 
algorithm \cite{Cacciari:2008gp,Cacciari200657}, with a radius parameter $R=0.4$,
using clusters of adjacent calorimeter cells calibrated at the electromagnetic (EM) energy scale.
These jets are corrected for the calorimeter response and other detector effects using energy and pseudorapidity dependent calibration factors derived from simulation and validated using data \cite{Aad:2011he}.
The calibrated jets, $j$, used in the analysis are required to have $p_{\rm T} > 25$~GeV, $|y|<2.4$ and are required to be well separated from the selected leptons $\ell$~(electrons or muons) by
\begin{equation}
\Delta R(j,\ell) = \sqrt{ \left(\Delta \phi(j ,\ell)\right)^2 + \left(\Delta \eta(j,\ell) \right)^2} > 0.4.
\end{equation}
Jets originating from $b$-quarks ($b$-jets) are identified using the IP3D+SV1 algorithm \cite{ATLAS-CONF-2011-102} and are referred to as $b$-tagged jets. This algorithm, based on impact parameter and secondary vertex information, has an average per-jet efficiency of $70\%$ for jets originating from $b$-quarks in simulated $t\bar{t}$ events and rejects approximately $99\%$~of
jets originating from light quarks and gluons.

The scalar sum of visible transverse momentum, $H_{\rm T}$, is calculated using the transverse momenta of all the reconstructed jets and leptons that satisfy the selection criteria defined above.
The missing transverse momentum, $\MET$, is reconstructed from EM-scale clusters corrected according to the energy scale of associated jets/electrons and the measured muon momenta.

To create a highly enriched $t\bar{t}$ sample, events are required to have two opposite sign high-$\pt$~leptons and at least two $b$-tagged jets. The analysis is then divided into the three dilepton decay channels, $ee$, $e\mu$ and $\mu\mu$, and additional channel-dependent selection criteria are applied to reduce backgrounds further.
The background in the $ee$ and $\mu\mu$ channels arising from $Z\to ee / \mu\mu$~events is suppressed by requiring $\MET > 40$~GeV and that the dilepton mass, $m_{\ell\ell}$, is not in the range of the $Z$-boson mass, i.e. $\left| m_{\ell\ell}-91{\text{ GeV}}\right| > 10$ GeV.
In addition, events are required to have $m_{\ell\ell} > 15$~GeV in order to reject backgrounds from vector-meson decays.
The backgrounds in the $e\mu$ channel from $Z\to\tau\tau$~and diboson events are suppressed by requiring $H_{\rm T}$ to be greater than 130 GeV.
A summary of the event selection criteria is presented in Table \ref{t:eventcuts}.

\begin{table*}
 \caption{Selection requirements applied to the three analysis channels.}
\begin{center}
\begin{tabular}{lccc}
\midrule
 & \multicolumn{3}{c}{Channel} \\
\cmidrule(l){2-4}
  Selection \hspace{0.3cm} & $ee$ & $\mu\mu$ & $e\mu$ \\[2pt]
\toprule
 \multirow{2}{*}{Electrons} & 2 with $E_{\rm T} > 25$~GeV,  &  \multirow{2}{*}{$-$} & 1 with $E_{\rm T} > 25$~GeV,  \\
                   & $|\eta|<2.47$ & & $|\eta|<2.47$ \\[3pt]
%\midrule
 \multirow{2}{*}{Muons}      &  \multirow{2}{*}{$-$} & 2 with $\pt > 20$~GeV,  & 1 with $\pt > 20$~GeV, \\
                   &        & $|\eta|<2.5$ & $|\eta|<2.5$ \\[3pt]
%\midrule
 $\MET$ & $> 40$~GeV & $> 40$~GeV & $-$ \\[3pt]
%\midrule
 $H_{T}$ & $-$ & $-$ & $> 130$~GeV \\[3pt]
%\midrule
  \multirow{2}{*}{$m_{\ell\ell}$} & $> 15$~GeV  & $> 15$~GeV &  \multirow{2}{*}{$-$} \\[1pt]
  & $\left| m_{\ell\ell}-91\, \rm{GeV}\right| > 10$~GeV &  \hspace{0.4cm}$\left| m_{\ell\ell}-91\, \rm{GeV}\right| > 10$~GeV\hspace{0.4cm} & \\[3pt]
 \cmidrule(l){2-4}
 $b$-tagged jets           & \multicolumn{3}{c}{ At least 2 with $\pt > 25$~GeV, $|y|<2.4$, $\Delta R(j, \ell) > 0.4$} \\
  \bottomrule
  \end{tabular}
  \end{center}
  \label{t:eventcuts}
  \end{table*}
 
\begin{figure*}[htbp]
 \begin{center}
 \subfigure[] {
 \includegraphics[width=0.45\textwidth]{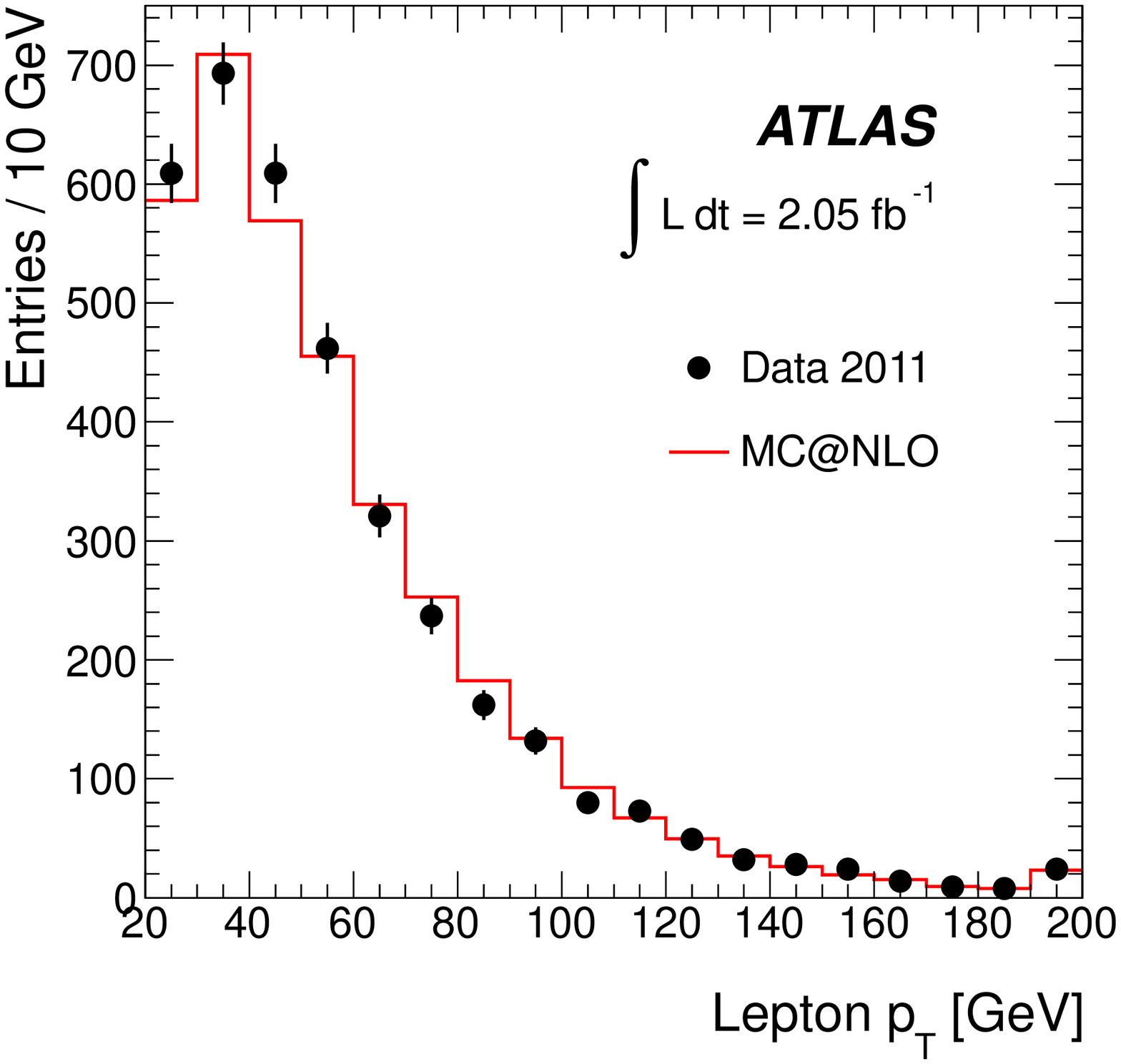}
 }
  \subfigure[] {
 \includegraphics[width=0.45\textwidth]{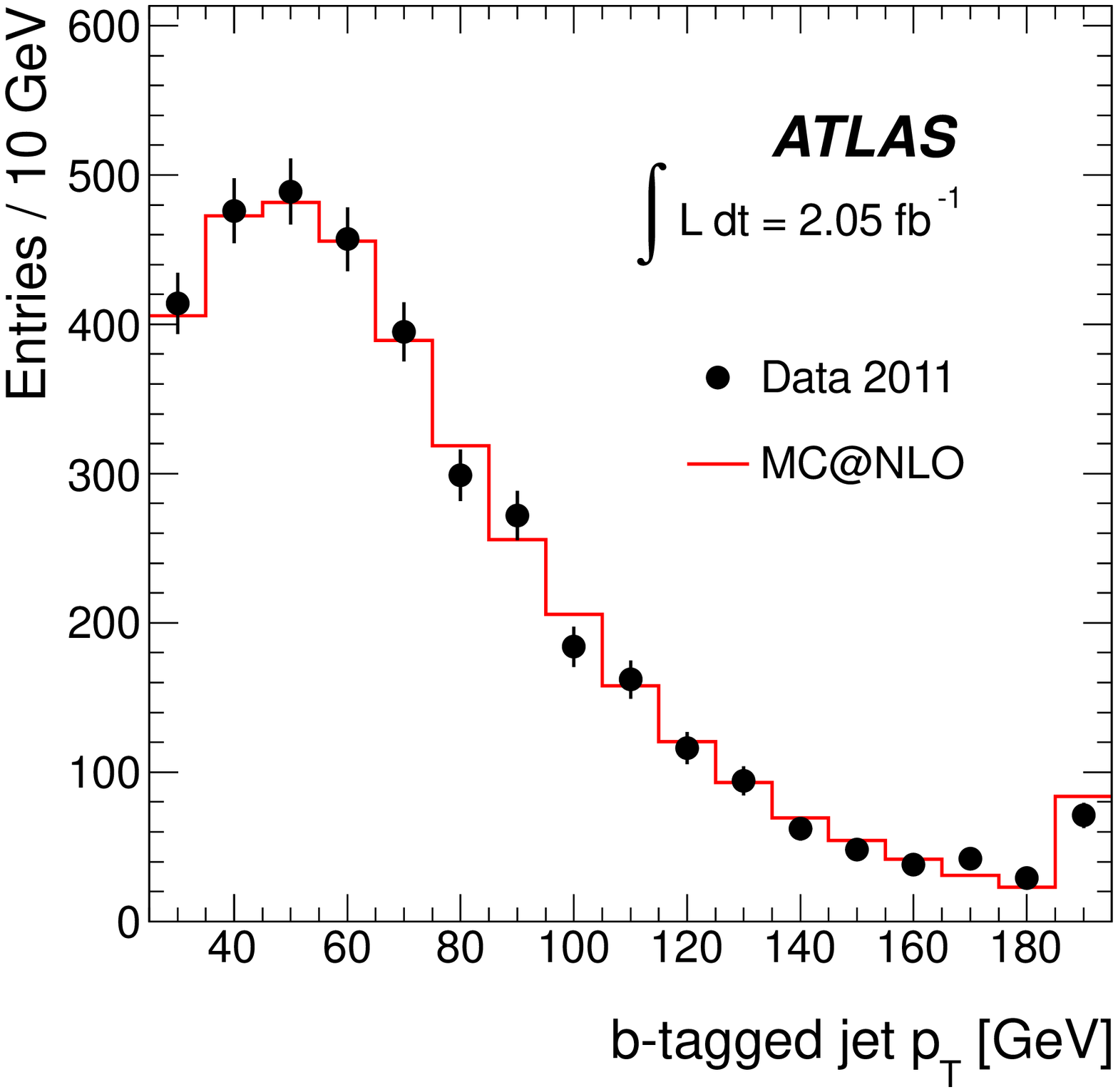}
 }
 \end{center}
 \caption{
 The distribution of (a) lepton $\pt$~and (b) $b$-tagged jet $\pt$~for the selected events compared to the \mcatnlo{} simulation of $t\bar{t}$ events. The data is shown as closed (black) circles with the statistical uncertainty. The \mcatnlo{} prediction is normalised to the data and is shown as a solid (red) line. The overflow events at high $p_{\rm T}$ are added into the final bin of each histogram.
 }
\label{f:controlplots}
 \end{figure*}

The number of selected events in the three channels is $242$~($ee$), $436$~($\mu\mu$) and $1095$~($e\mu$).
The dominant background contributions after the selection requirements are single top ($Wt$) production and events in which at least one lepton originates from heavy flavour decay or jet misidentification. The latter contribution consists of mainly $W$+jets and multijet events and is estimated from the data
using a method described in reference \cite{inreview1}. The $Wt$~background is estimated
using the MC sample discussed in Section~\ref{sec-mcsamples}. The total background contamination is estimated to be less than 6\%, which is smaller than the uncertainty on the theoretical calculation of the $t\bar{t}$ cross section \cite{Moch:2008qy,Langenfeld:2009tc,Beneke:2009ye}. 
The expected background contributions are not subtracted from the data, but are considered as a source of systematic uncertainty on the measurement.
Figure~\ref{f:controlplots} shows the distribution of the lepton and $b$-tagged jet $\pt$~for the selected data events
compared with the prediction from the \mcatnlo{} $\ttbar$ simulation.
Good agreement is seen in all such distributions.

\begin{figure*}[htbp]
 \begin{center}
 \subfigure[] {
 \includegraphics[width=0.45\textwidth]{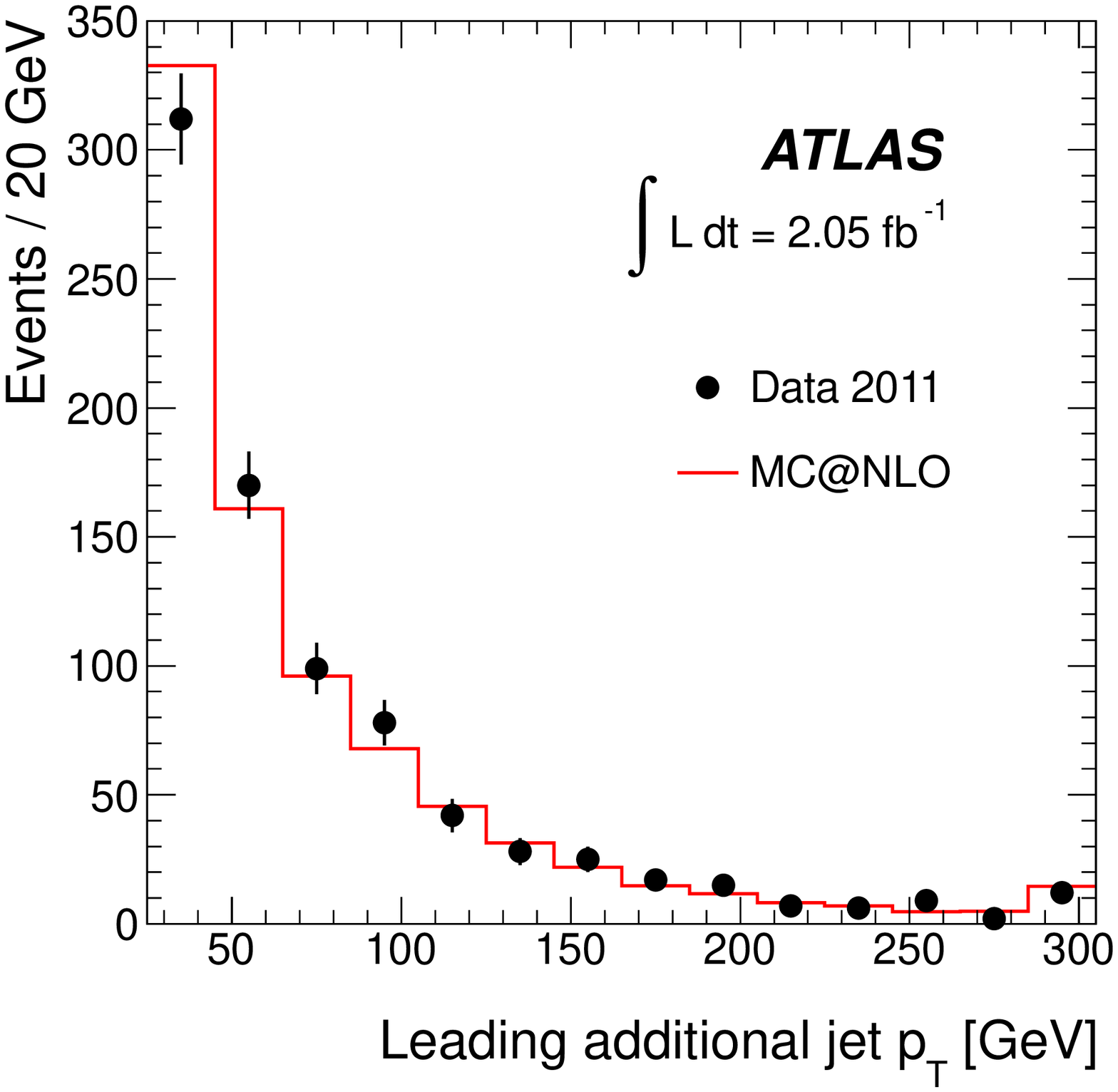}
 }
  \subfigure[] {
 \includegraphics[width=0.45\textwidth]{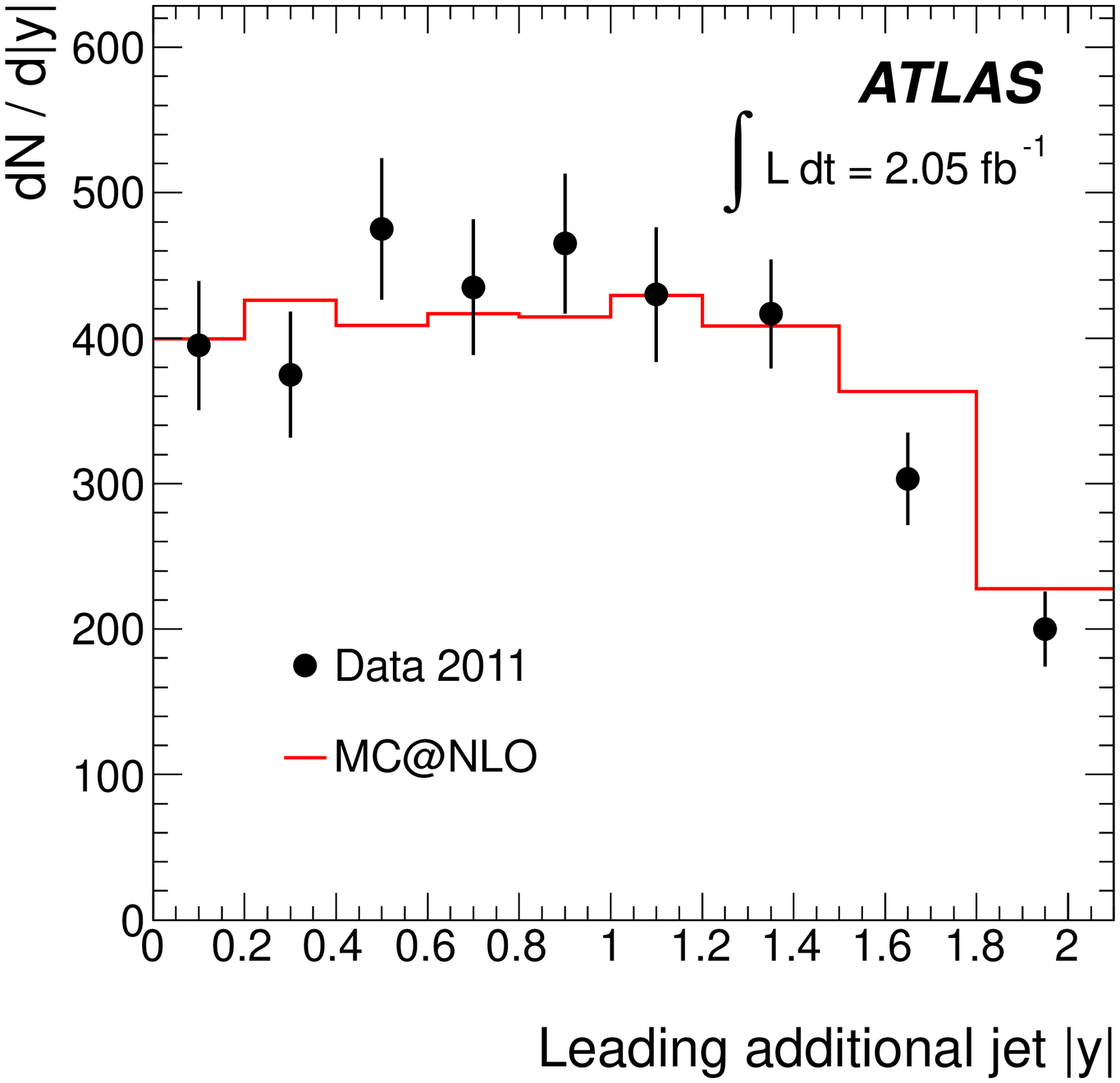}
 }
 \end{center}
 \caption{
 Distribution of (a) leading additional jet $\pt$~and (b) leading additional jet rapidity in the selected events compared to the \mcatnlo{} simulation of $t\bar{t}$ events.  The data is shown as closed (black) circles with the statistical uncertainty. The \mcatnlo{} prediction is normalised to the data and is shown as a solid (red) line. In the $\pt$~distribution, the overflow events at high $p_{\rm T}$ are added into the final bin of the histogram. In the rapidity distribution, variable bin sizes are used such that the bin edges  match the rapidity intervals used to construct the gap fractions.
 }
\label{f:addrecojet}
 \end{figure*}

The gap fraction in each rapidity interval is computed using the additional jets in the event. To suppress jets from overlapping proton-proton collisions, the additional jets are required to be fully contained within the inner detector acceptance ($|y|<2.1$) and the jet vertex fraction (JVF) algorithm is used to identify jets from the primary interaction. After associating tracks to jets ($\Delta R(\rm jet, track) < 0.4$), the JVF is defined as the scalar summed transverse momentum of associated tracks from the primary vertex divided by the summed transverse momentum of associated tracks from all vertices. Each additional jet is required to satisfy ${\mathrm{JVF}} > 0.75$.
The transverse momentum and rapidity distributions for the highest-$p_{\rm T}$ additional jet in the region $|y|<2.1$ 
is shown in Figure~\ref{f:addrecojet}.
Reasonable agreement is seen between the data and the  \mcatnlo{} $\ttbar$ simulation.
 
\section{Correction for Detector Effects}
\label{sec-unfolding}

The data are corrected for detector effects to produce results at the particle level. 
The particle level $t\bar{t}$ events are defined in each channel using the same event selection criteria applied to the reconstructed data, as presented in Table \ref{t:eventcuts}.  Final state stable particles are defined as those that have a mean lifetime $c\tau >10$~mm. Electrons are required to have $E_{\rm T} > 25$~GeV and $|\eta|<2.47$, whereas 
muons are required to have $p_{\rm T} > 20$~GeV and $|\eta|<2.5$\footnote{Changing the muon selection criteria to match the electron fiducial region ($p_{\rm T}>25$~GeV and $|\eta|<2.47$) was observed to have a negligible impact on the gap fraction.}. Jets are reconstructed using the anti-$k_{\rm t}$ algorithm with $R=0.4$, using all stable particles except muons and neutrinos, and are required to have $p_{\rm T}>25$~GeV and $|y| < 2.4$. Jets originating from $b$-quarks are defined as any jet that is within $\Delta R<0.3$ of a $B$-hadron, where the $B$-hadrons are required to have $\pt>5$~GeV. $H_{\rm T}$ is defined as the scalar sum of jet and lepton transverse momenta and $\met$ is defined using all final state neutrinos.

The correction factor, $C$, for the gap fraction at a specific value of $x=\Qz$~or~$\Qsum$, is defined as
\begin{equation}
C(x) = \frac{f^{\rm truth}(x)}{f^{\rm reco}(x)},
\end{equation}
where $f^{\rm reco}(x)$~is the reconstructed gap fraction and $f^{\rm truth}(x)$~is the particle level gap fraction. 
The use of simple correction factors is justified because the purity of the selected events is greater than 70\% for each value of $\Qz$ or $\Qsum$. The purity of the selected events is defined as the number of events that pass the event selection at both the reconstructed and particle level, divided by the number of events that pass the event selection at reconstructed level, using the \mcatnlo{} simulation of $t\bar{t}$ events.

The \mcatnlo{} simulation is also used to derive the baseline correction factors used in this measurement. These correction factors depend on the rapidity interval used to veto jet activity, with corrections of $2\%-5\%$ for $\Qz=25$ GeV that decrease with increasing $\Qz$. The systematic uncertainties on these correction factors due to physics and detector modelling are discussed in Section~\ref{sec-syst}.

\section{Systematic Uncertainties}
\label{sec-syst}

\begin{figure*}[ht]
\begin{center}
\subfigure[]{
	\includegraphics[width=0.47\textwidth]{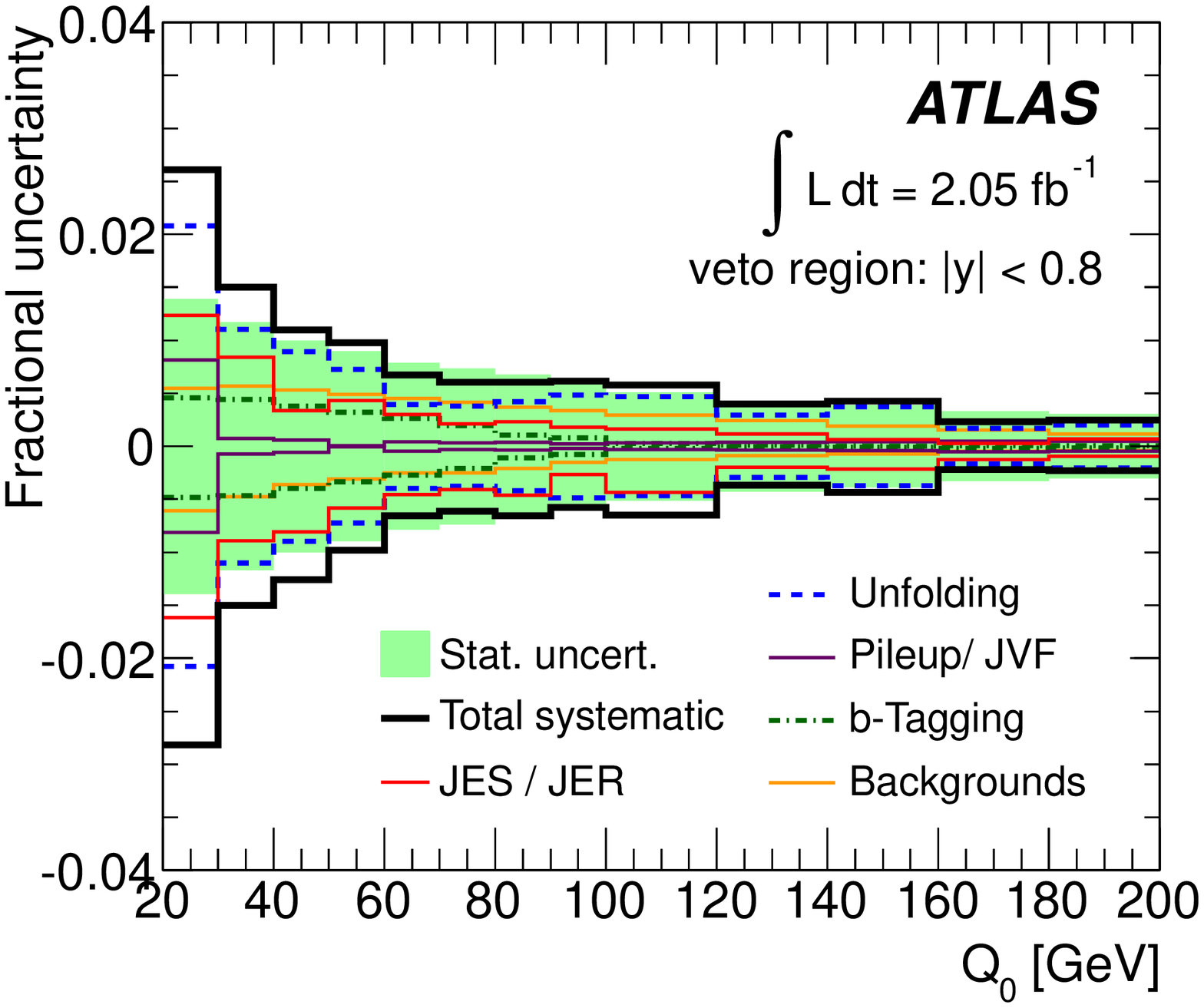}
}
\subfigure[]{
	\includegraphics[width=0.47\textwidth]{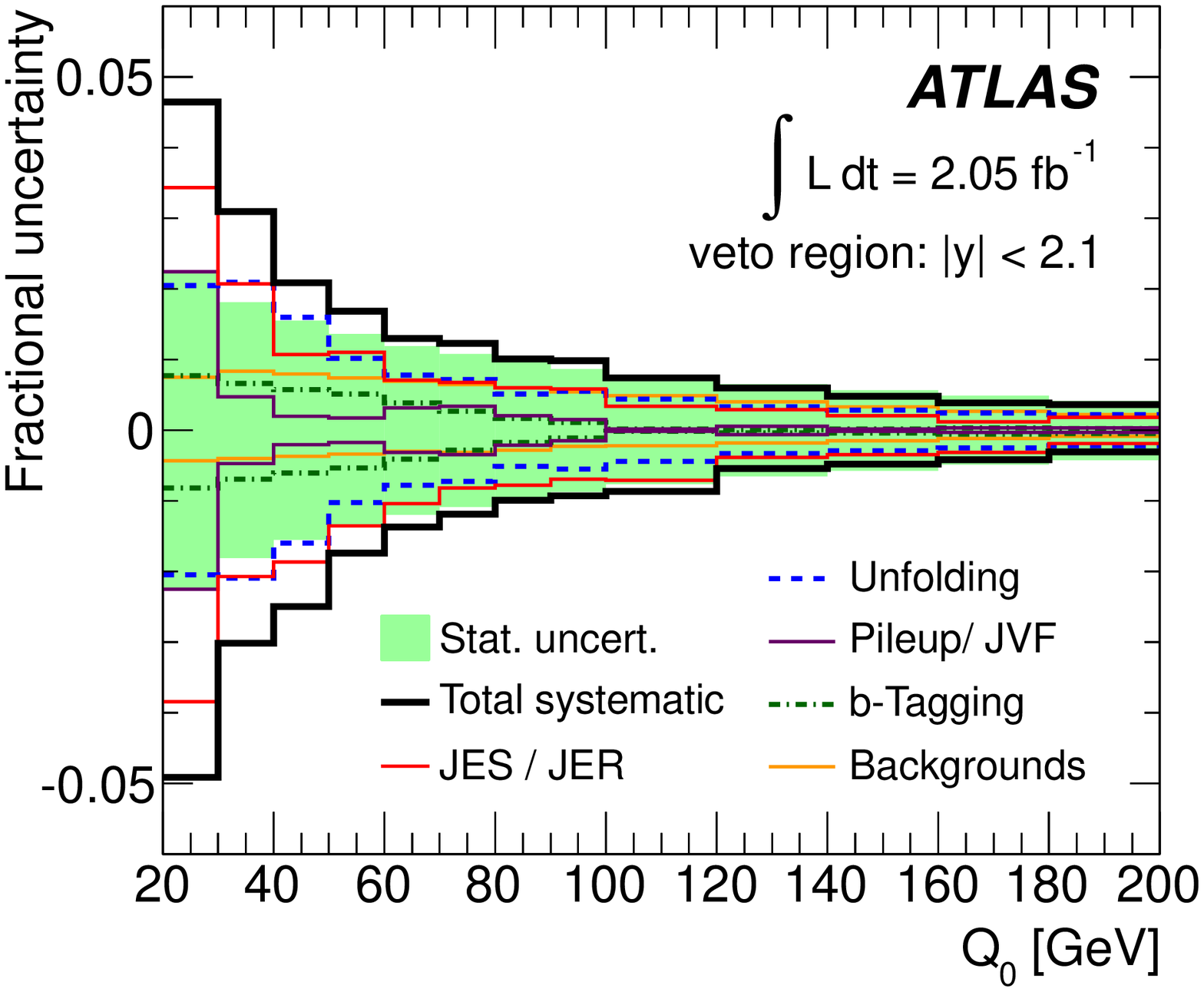}
}
\end{center}
\caption{
Breakdown of the systematic uncertainties on the gap fraction as a function of $\Qz$~for (a) $|y|<0.8$ and (b) $|y|<2.1$. The step size in $\Qz$ was chosen to be commensurate with the jet energy resolution. The individual systematic uncertainties are shown as labelled lines of different styles and the total systematic uncertainty is shown as the outer solid line. The statistical uncertainty on the data is shown as the shaded area. The breakdown of the systematic uncertainties above $\Qz=200$~GeV is consistent with the results at $\Qz=200$~GeV. `Pileup' refers to the effect of jets produced in a different proton-proton interaction. `Unfolding' refers to the procedure used to correct the measured gap fraction to particle level.
}
\label{f:syst:summary}
\end{figure*}

Uncertainties related to the inclusive $\ttbar$~event selection were found to cancel in the gap fraction and are neglected in the final systematic uncertainty. These include the uncertainties on the lepton momentum scale, momentum resolution and reconstruction efficiency, the $b$-jet energy scale, the trigger efficiency for each analysis channel and the integrated luminosity. The dominant sources of systematic uncertainty are those that directly affect the additional jets. These non-negligible sources of uncertainty are discussed in this section and a summary is presented in Figure \ref{f:syst:summary}.

The experimental aspects that affect the additional jets are the jet energy scale (JES), the jet energy
resolution (JER), the jet reconstruction efficiency and the JVF selection requirement. The uncertainty on the gap fraction due to the JES is estimated by rescaling the jet energies in the simulation by the known uncertainty  \cite{Aad:2011he}. The uncertainty on the JES includes the impact of soft energy added to jets from multiple proton-proton interactions.
The uncertainty on the gap fraction due to jet reconstruction efficiency \cite{Aad:2011he} and the jet energy resolution  is estimated by varying each of these in the simulation within the allowed uncertainties determined from data. 
The relative uncertainty on the gap fraction due to the JES and JER uncertainties is $3.5\%$ or less if jets are vetoed in the full rapidity interval ($|y|<2.1$), and $1.5\%$ or less if jets are vetoed in the smaller sub-intervals (e.g. $|y|<0.8$).
The uncertainty from the jet reconstruction efficiency is found to be negligible compared to the JES and JER uncertainties for all four rapidity intervals.

 The bias due to the JVF selection efficiency is estimated by performing the full analysis (selection plus correction for detector effects) with a relaxed requirement of ${\mathrm{JVF}} > 0.1$.
The relative difference between the results obtained with the standard and relaxed  requirement is found to be up to $2\%$~at $\Qz=25$~GeV and is negligible above $\Qz$~of approximately $100$~GeV.
This difference is taken as the systematic uncertainty due to the JVF selection efficiency.

Jets produced by additional proton-proton interactions are suppressed by the JVF requirement. However, those jets that pass this requirement represent a potential bias in the measurement.
The size of this bias is evaluated by removing those jets in the \mcatnlo{} sample that are not matched
to a particle level jet from the $pp$~interaction that produces the $\ttbar$~event. The matching criterion is $\Delta R<0.3$ and the particle jet transverse momentum is allowed to be as low as 7~GeV, to avoid resolution effects in the matching procedure. The gap fraction is recalculated using this truth-matched sample and the difference to the nominal gap fraction is taken as the systematic uncertainty due to jets from additional proton-proton interactions. 
The relative uncertainty on the gap fraction is less than $1\%$ in each of the rapidity regions.

Background contamination is treated as a systematic uncertainty. For each background source, the expected events are subtracted from the data and the gap fraction is re-calculated. The relative difference with respect to the nominal result is taken as the systematic uncertainty due to background contamination; the largest effect is observed to be 0.5\% for $\Qz=25$~GeV.

The uncertainty on the efficiency and rejection capability of the $b$-tagging algorithm impacts upon the measurement if the additional jet is identified as a $b$-tagged jet instead of one of the $b$-jets originating from the top-quark decay. The systematic uncertainty due to this effect is estimated by changing the baseline efficiency and rejection corrections, which are applied to the simulation, according to the $b$-tagging uncertainty (derived in calibration studies using inclusive lepton and multijet final states). The relative uncertainty on the gap fraction is less than $0.8\%$.

The uncertainty on the procedure used to correct the data to particle level due to physics modelling is
estimated by deriving alternative correction factors using the \powheg{} samples.
The systematic uncertainty in the correction procedure is taken to be the largest difference between the correction factor obtained using the \mcatnlo{} sample and the correction factor obtained using the two \powheg{} samples. In the case where this difference is smaller than the statistical uncertainty in the MC samples, the statistical uncertainty is taken as the estimate of the systematic uncertainty.
The relative uncertainty on the correction factors is less than 2\% at $\Qz=25$~GeV for the region $|y|<2.1$, decreasing to
approximately $0.3\%$~at $\Qz=150$~GeV. The sensitivity of the corrections to the physics modelling is further assessed by reweighting the additional jet $\pt$~spectrum in the \mcatnlo{} sample such that the $\pt$~distribution has the maximal change in shape that is consistent with the JES uncertainty bands. The difference in the correction factors was observed to be much smaller than the differences obtained by using different MC generators and is neglected in the final results.

Figure~\ref{f:syst:summary} shows the breakdown of the systematic uncertainties on the gap fraction
as a function of $\Qz$, for the veto regions $|y|<0.8$ and $|y|<2.1$.
This Figure also shows the total systematic uncertainty, which is calculated by adding in quadrature all the individual systematic uncertainties.
The total systematic uncertainty is largest at low $\Qz$ and is dominated by the jet related uncertainties (JES, JER and JVF) and the uncertainty on the correction factors. The measurement is most precise in the central region, where the jet energy scale uncertainty is smallest. 
The breakdown of uncertainties for the gap fraction as a function of $\Qsum$~is similar, but the uncertainties are slightly larger and fall more slowly as a function of $\Qsum$.
This is due to low transverse momentum jets, which have the largest systematic uncertainties and therefore affect all values of $\Qsum$.

\section{Results and Discussion}
\label{sec-results}

The gap fraction is measured for multiple values of $\Qz$ and $\Qsum$ in the four rapidity intervals defined in Section~\ref{sec-intro}.
The step size in $\Qz$~and $\Qsum$~was chosen to be commensurate with the jet energy resolution. 
The results are corrected to the particle level as described in Section~\ref{sec-unfolding}. 

 \begin{figure*}[htbp]
 \begin{center}
 \subfigure[] {
 	\includegraphics[width=0.48\textwidth]{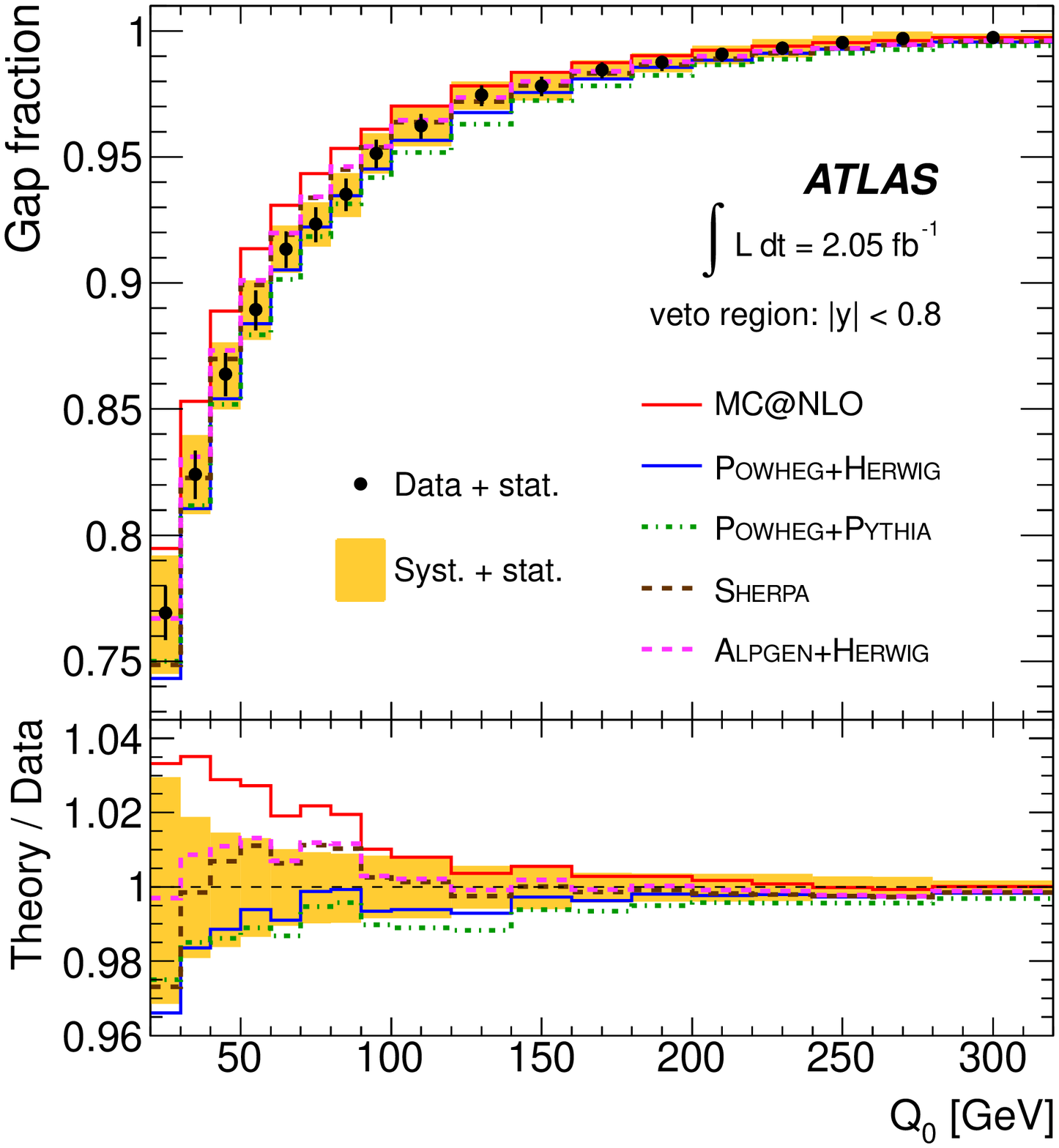}
	\label{f:unfoldedgf:nlo:central}
 } 
  \subfigure[] {
 	\includegraphics[width=0.48\textwidth]{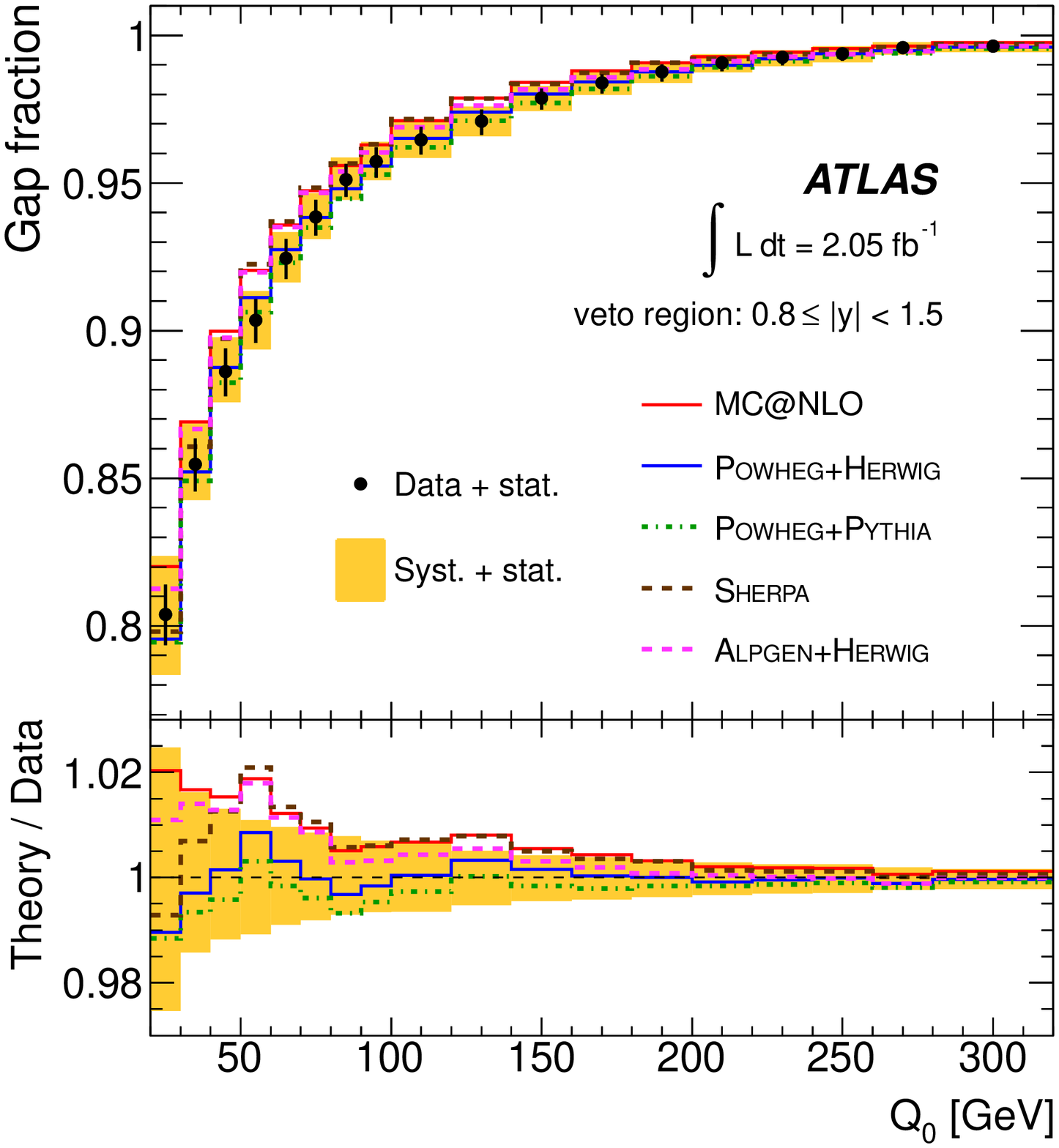}
 }
  \subfigure[] {
 	\includegraphics[width=0.48\textwidth]{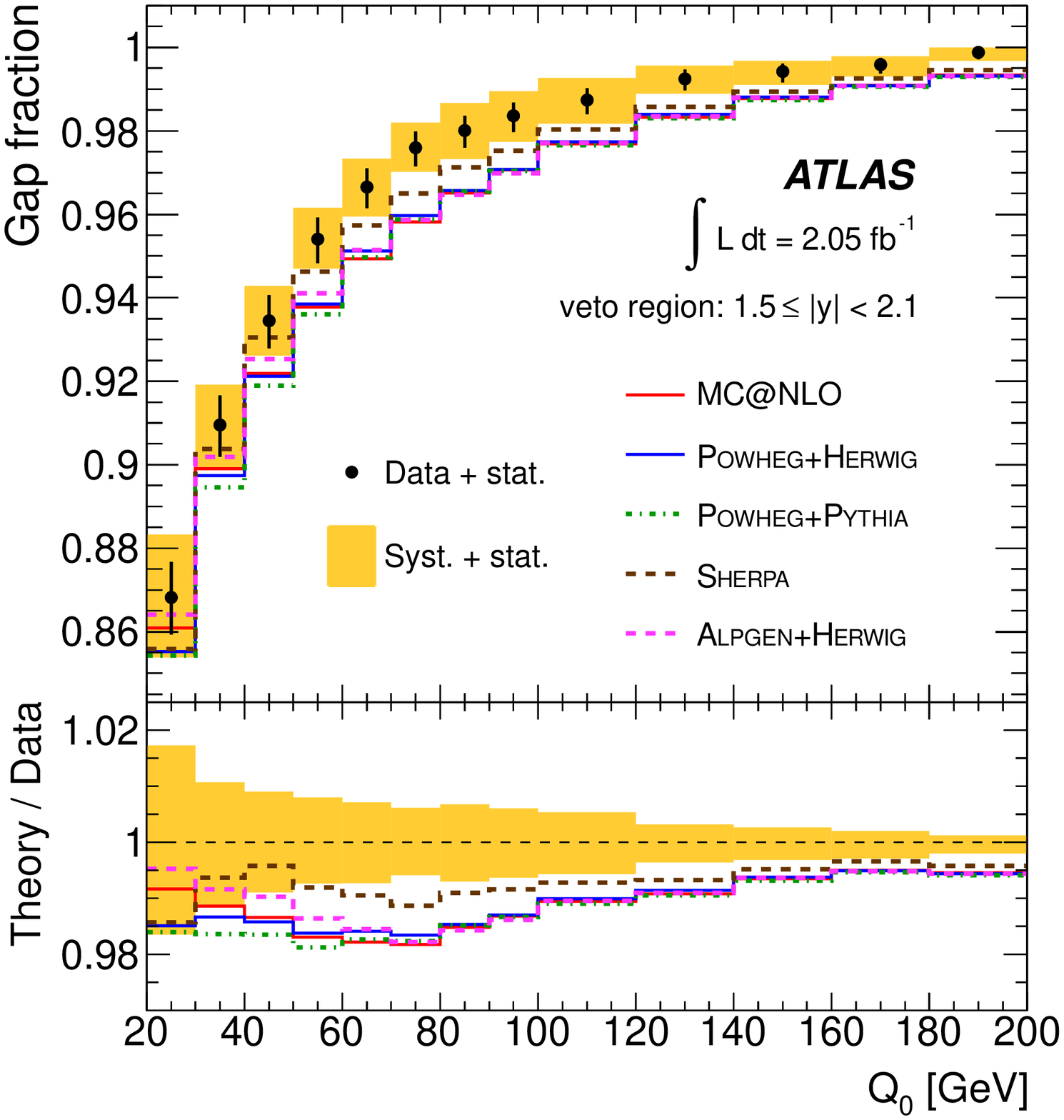}
	\label{f:unfoldedgf:nlo:fwd}
 }
  \subfigure[] {
 	\includegraphics[width=0.48\textwidth]{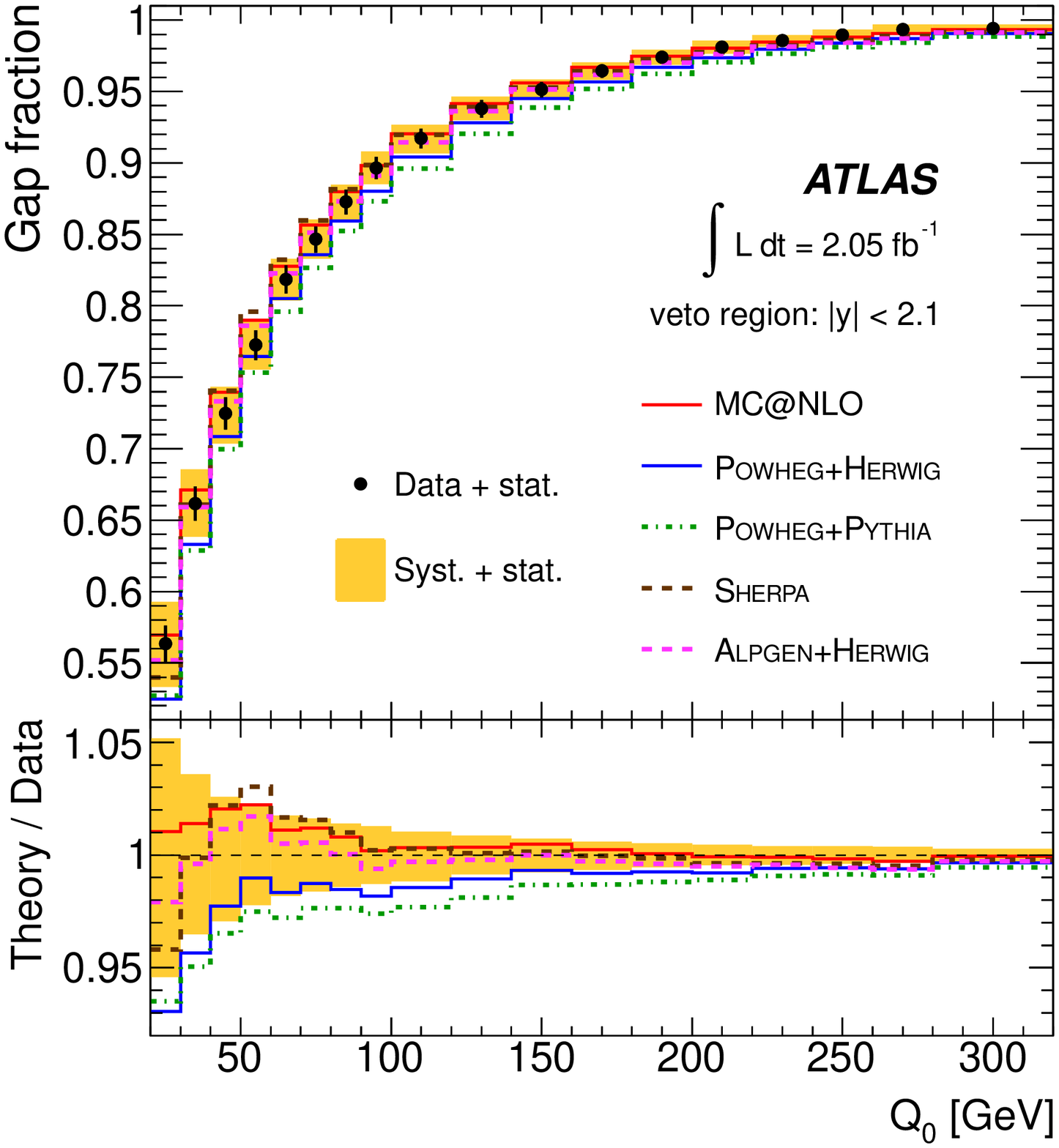}
	\label{f:unfoldedgf:nlo:all}
 }
 \caption{
 The measured gap fraction as a function of $\Qz$~is compared with the prediction
 from the NLO and multi-leg LO MC generators in the three rapidity regions, (a) $|y|<0.8$, (b) $0.8 \leq |y| < 1.5$ and
 (c) $1.5 \leq |y| < 2.1$. Also shown, (d), is the gap fraction for the full rapidity range $|y| < 2.1$. The data is represented as closed (black) circles with statistical uncertainties. The yellow band is the total experimental uncertainty on the data (statistical and systematic). The theoretical predictions are shown as solid and dashed coloured lines. The gap fraction is shown until $\Qz=300$~GeV or until the gap fraction reaches one if that occurs before $\Qz=300$~GeV.
 }
 \label{f:unfoldedgf:nlo}
 \end{center}
 \end{figure*}

\begin{figure*}[htbp]
 \begin{center}
 \subfigure[] {
 	\includegraphics[width=0.48\textwidth]{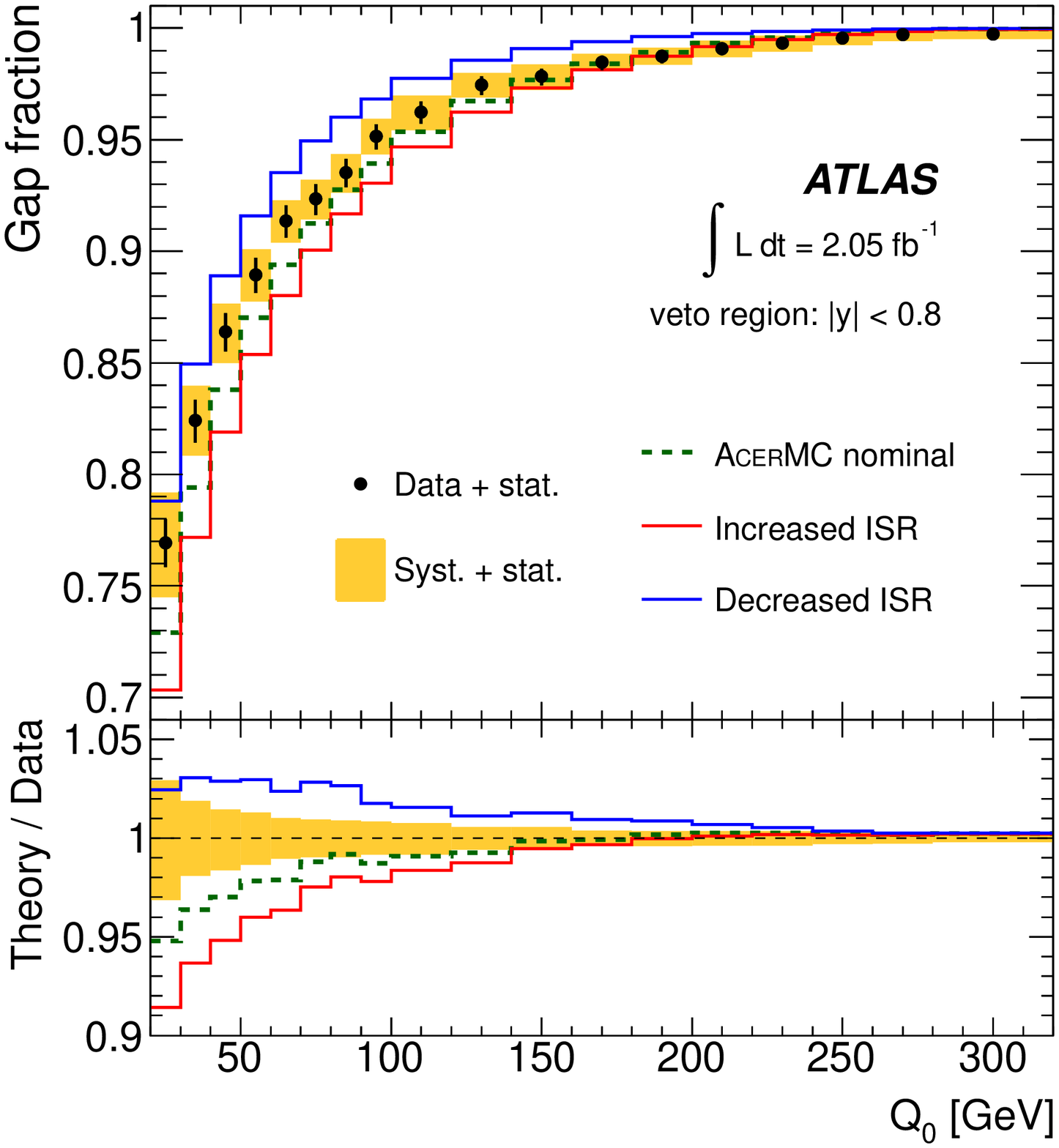}
 }
  \subfigure[] {
 	\includegraphics[width=0.48\textwidth]{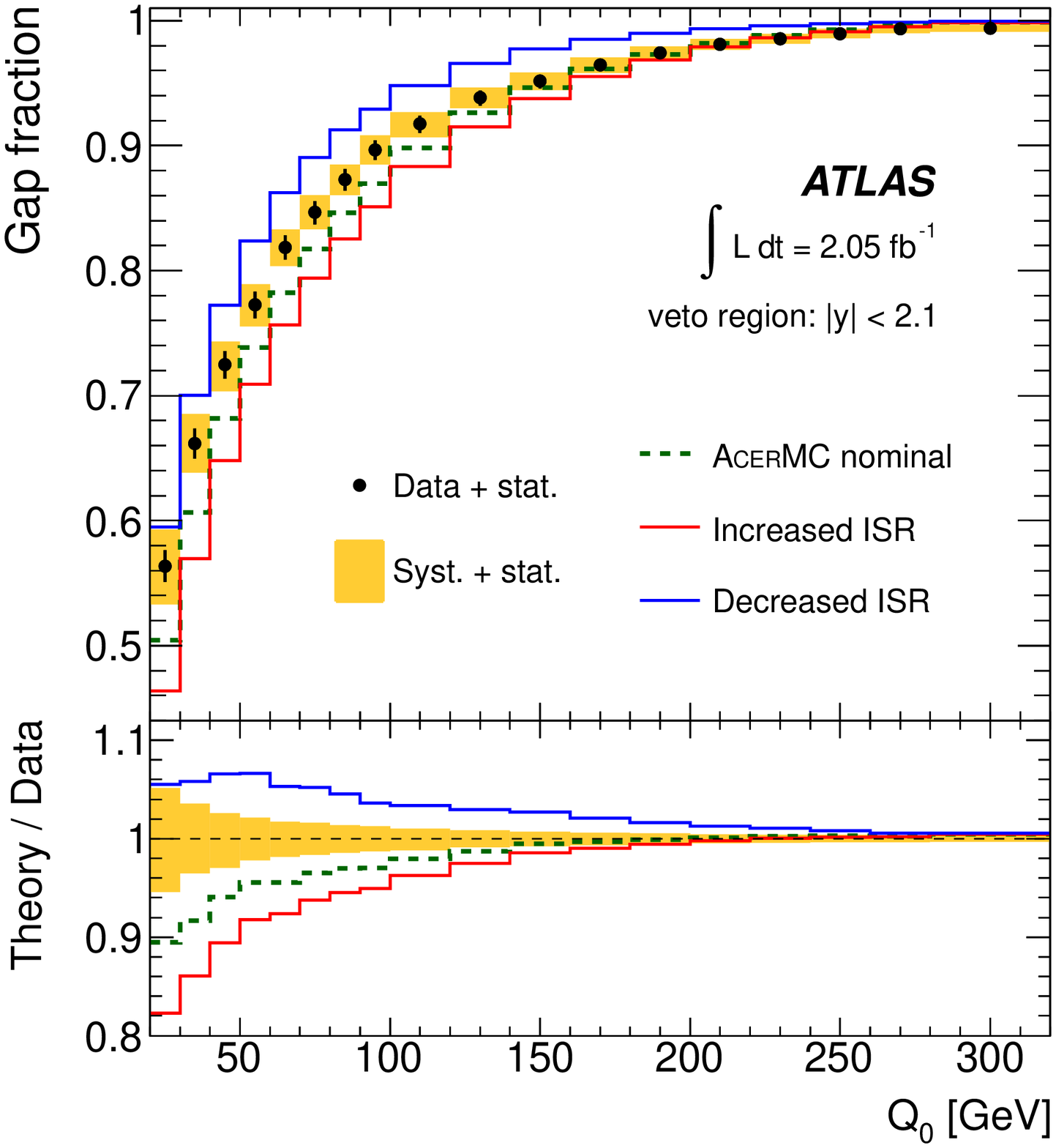}
 }
 \caption{
 The measured gap fraction as a function of $\Qz$~for (a) $|y|<0.8$~and (b) $|y| < 2.1$~is compared with the prediction
 from the \acermc~generator, where different settings of the \pythia~parton shower parameters are used
 to produce samples with nominal, increased and decreased initial state radiation (ISR).
The data and theory predictions are represented in the same way as in Figure \ref{f:unfoldedgf:nlo}.
 }
 \label{f:unfoldergf:acer}
 \end{center}
 \end{figure*}

The measured gap fraction as a function of $\Qz$~is compared with the predictions from the multi-leg LO and NLO generators in Figure~\ref{f:unfoldedgf:nlo}.
In general, all these generators are found to give a reasonable description of the data if the veto is applied to jets in the full rapidity interval, $|y|<2.1$ (Figure~\ref{f:unfoldedgf:nlo:all}). 
The difference between the \mcatnlo{} and \powheg{} predictions is similar to the precision
achieved in the measurement and as such the measurement is probing the different
approaches to NLO plus parton-shower event generation.

In the most central rapidity interval, $|y|< 0.8$, the gap fraction predicted by \mcatnlo{} is too large (Figure~\ref{f:unfoldedgf:nlo:central}). 
The tendency of \mcatnlo{} to produce fewer jets  than \alpgen{} at central rapidity has been discussed in the literature~\cite{Mangano:2006rw} and the measurement presented here 
is sensitive to this difference. In the most forward rapidity interval, none of the predictions agrees with the data for all values of $\Qz$ (Figure~\ref{f:unfoldedgf:nlo:fwd}). In particular, although \mcatnlo{}, \powheg{}, \alpgen{} and \sherpa{} produce similar predictions, the gap fraction is too small, implying that too much jet activity is produced by these event generators in the forward rapidity region. 
 
The predictions from the \acermc~generator with the variations of the \pythia{} parton shower
parameters are compared to the data in Figure~\ref{f:unfoldergf:acer} and are found to be in poor agreement with the data. The spread
of the predicted gap fraction due to the parameter variations is found to be much
larger than the experimental uncertainty, indicating that the variations can
be significantly reduced in light of the measurement presented in this article.
 
\begin{figure*}[htbp]
 \begin{center}
 \subfigure[] {
 	\includegraphics[width=0.48\textwidth]{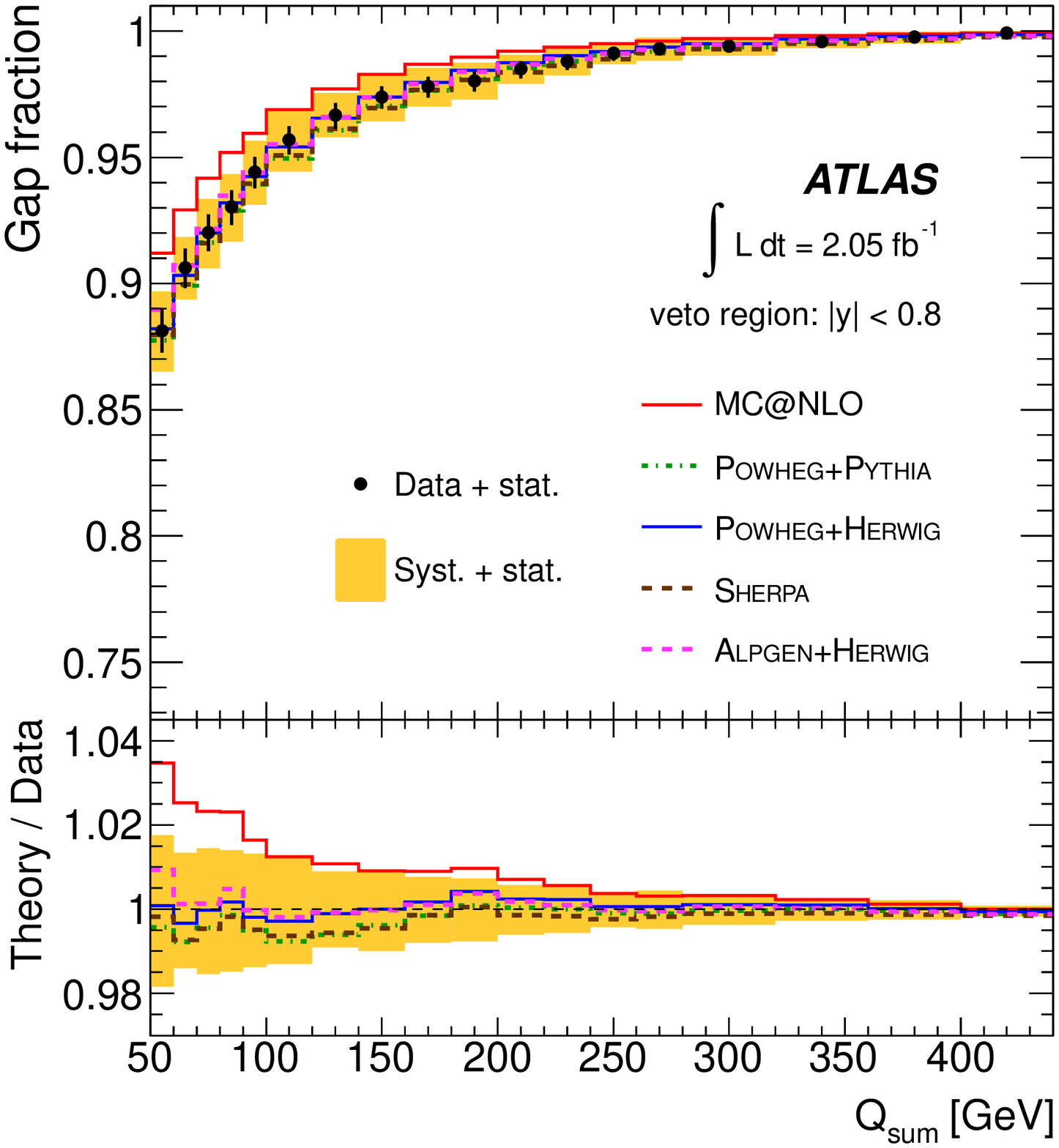}
 }
 \subfigure[] {
 	\includegraphics[width=0.48\textwidth]{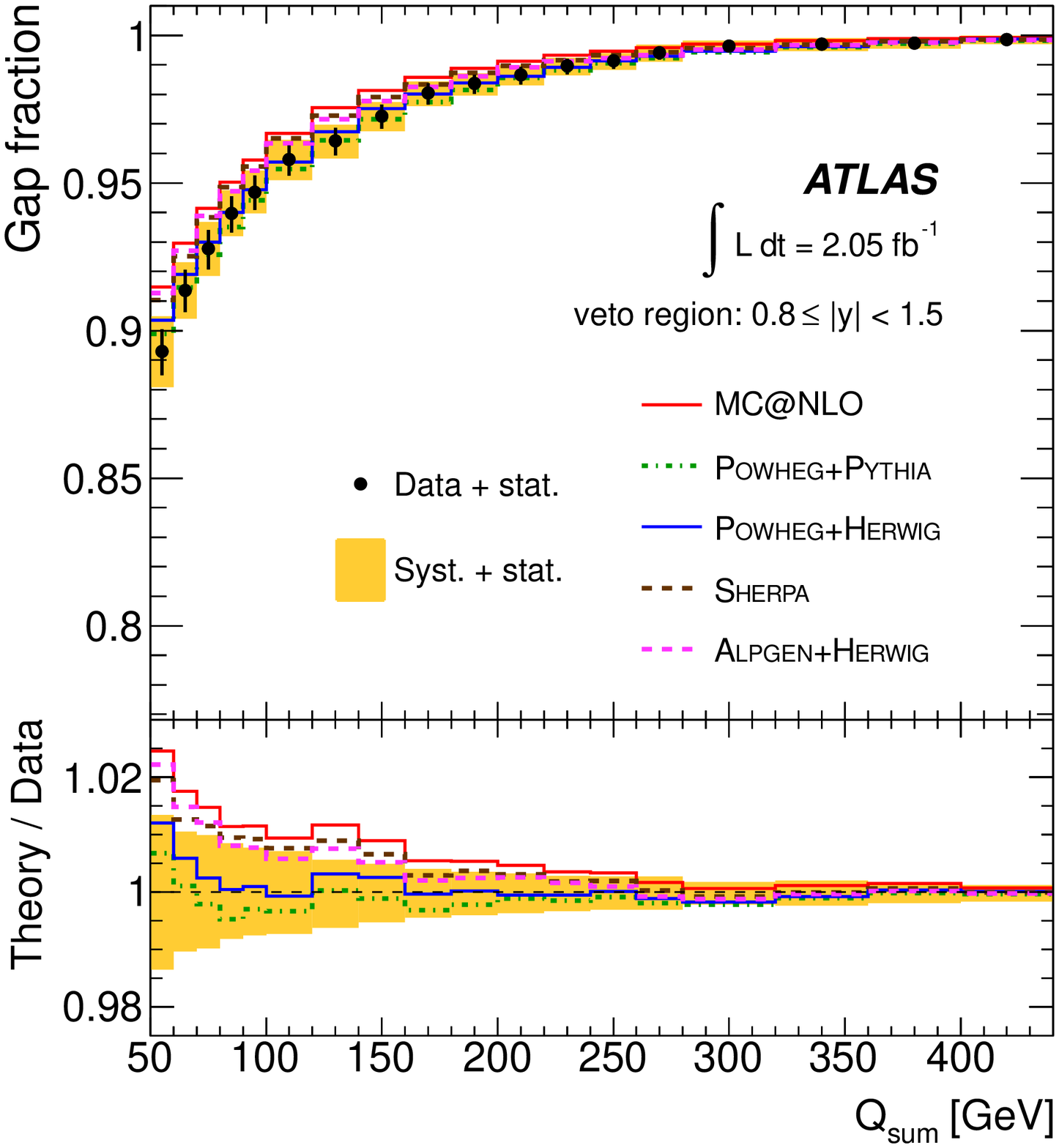}
 }
  \subfigure[] {
 	\includegraphics[width=0.48\textwidth]{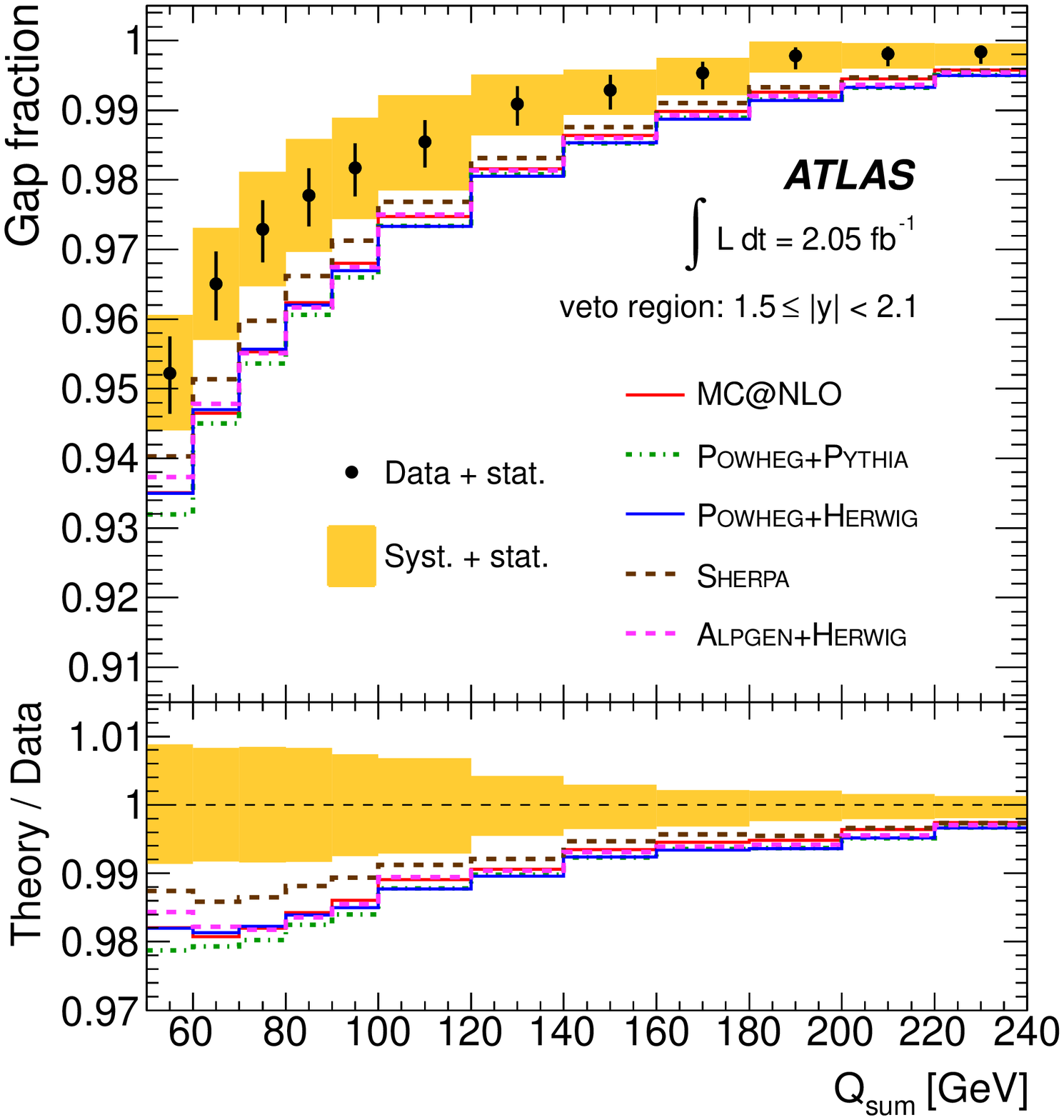}
 }
  \subfigure[] {
 	\includegraphics[width=0.48\textwidth]{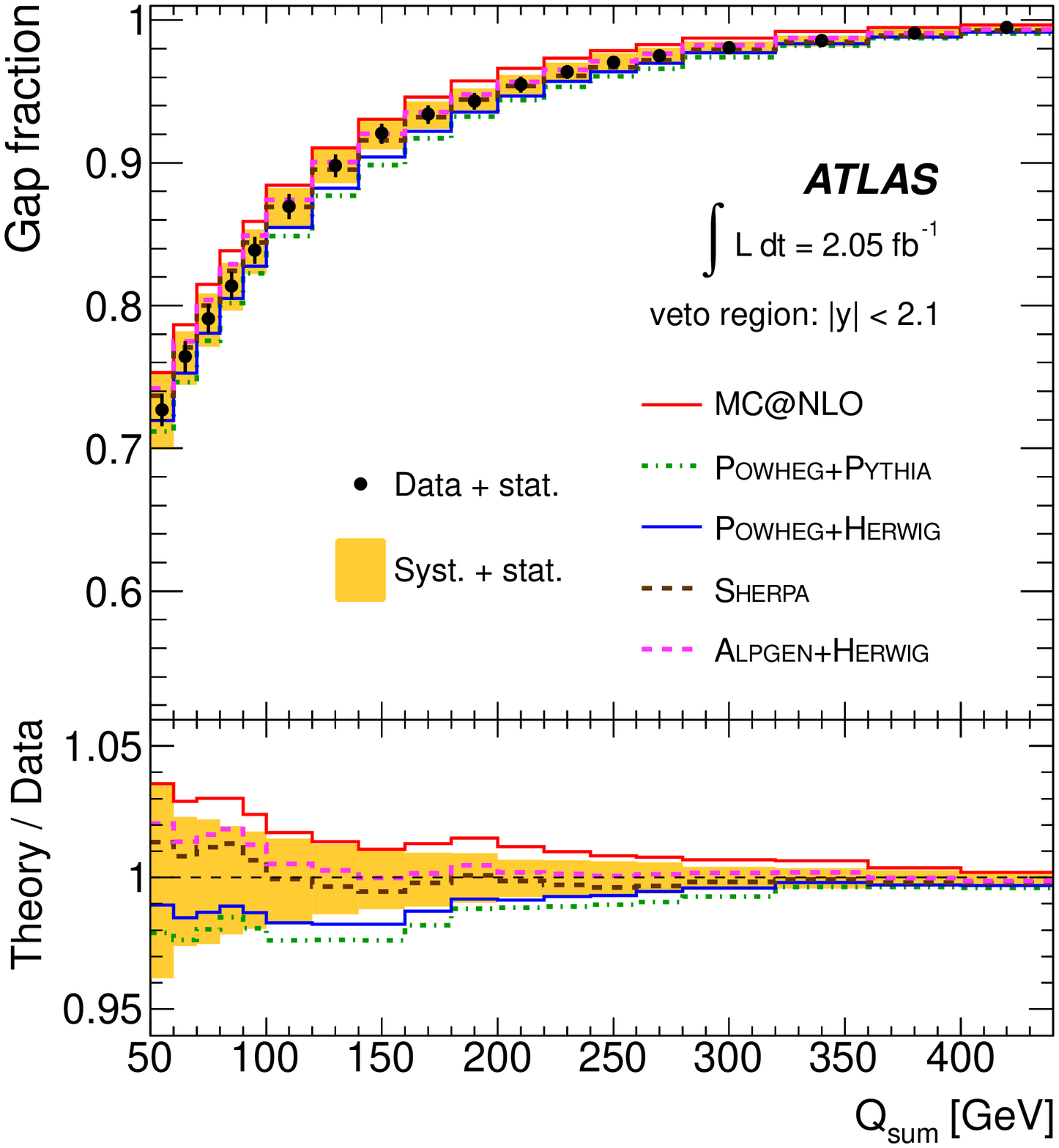}
 }
 \caption{
 The measured gap fraction as a function of $\Qsum$~is compared with the prediction
 from the NLO and multi-leg LO MC generators in the three rapidity regions, (a) $|y|<0.8$, (b) $0.8 \leq |y| < 1.5$ and
 (c) $1.5 \leq |y| < 2.1$. Also shown, (d), is the gap fraction for the full rapidity range $|y| < 2.1$. The data and theory predictions are represented in the same way as in Figure \ref{f:unfoldedgf:nlo}.
 The gap fraction is shown until $\Qsum=420$~GeV or until the gap fraction reaches one if that occurs before $\Qsum=420$~GeV.
 }
 \label{f:unfoldedsumgf:nlo}
 \end{center}
 \end{figure*}

The measured gap fraction as a function of $\Qsum$~is compared with the multi-leg LO and NLO generators in Figure~\ref{f:unfoldedsumgf:nlo}. The gap fraction is lower than for the case of the $\Qz$~variable, demonstrating that the measurement is probing quark and gluon radiation beyond the first emission. As expected, the largest change in the gap fraction occurs when jets are vetoed in the full rapidity interval, $|y|<2.1$. However, the difference between the data and each theoretical prediction is found to be similar to the $\Qz$~case. This implies that, for this variable, the parton shower approximations used for the subsequent emissions in \mcatnlo{} and \powheg{} are performing as well as the LO approximations used in \alpgen{} and \sherpa{}.

\begin{figure}[htbp]
 \begin{center}
 \subfigure[] {
 	\includegraphics[width=0.48\textwidth]{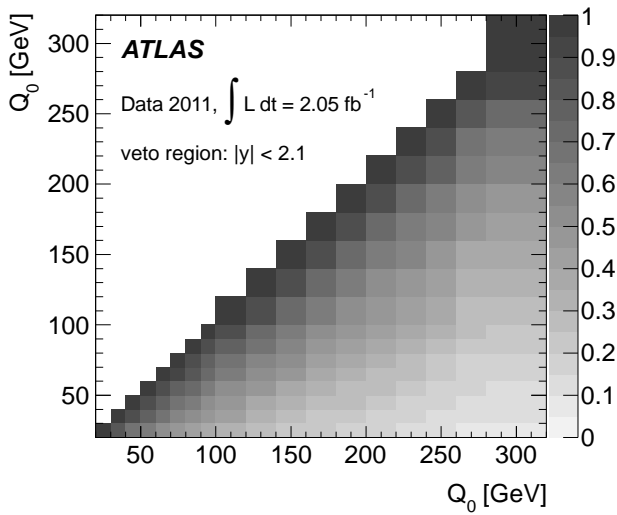}
 }
 \caption{
 The correlation matrix (statistical) for the gap fraction measurement at different values of $\Qz$ for $|y|<2.1$. 
 }
 \label{f:correlations}
 \end{center}
 \end{figure}

The gap fraction is a ratio of cross sections and all the events are used to evaluate this ratio at each value of $\Qz$~or $\Qsum$. This means that there is a statistical correlation between the measured
gap fraction values in each rapidity interval. 
The correlation matrix is shown in Figure~\ref{f:correlations} for the
gap fraction at different values of $\Qz$ for the $|y|<2.1$~rapidity region.
Neighbouring $\Qz$~points have a significant correlation, whereas well separated $\Qz$\ points are less correlated. 

The measured values of the gap fraction at $\Qz=25, 75$~and $150$~GeV are presented in Table \ref{t:sum:q0} for the different rapidity intervals used to veto jet activity. The statistical correlations between these measurements and the predictions from the multi-leg LO and NLO generators are also given. The measured values of the gap fraction at $\Qsum=55, 150$~and $300$~GeV are presented in Table \ref{t:sum:qsum} for the different rapidity intervals used to veto jet activity. The complete set of measurements presented in Figures \ref{f:unfoldedgf:nlo}-\ref{f:correlations} have been compiled in tables that can be obtained from HEPDATA.
 
 {
\renewcommand{\arraystretch}{1.27}
\begin{table*}
\caption{
The measured values of $f(\Qz)$~for $\Qz=25, 75$~and $150$~GeV for the different rapidity intervals used to veto jet activity are presented. The predictions from the NLO and multi-leg LO generators are also presented; the statistical uncertainty due to limited sample size is shown if this uncertainty is larger than 0.1\%. In each rapidity interval, the statistical correlations ($\rho^{\rm i}_{\rm j}$) between measurements at $\Qz=i$ and $\Qz=j$ are given. 
}\label{t:sum:q0}

\begin{center}
\begin{tabular}{lccccccc}
\midrule
\multicolumn{8}{c}{$f(\Qz)$ (\%)} \\
\cmidrule(l){2-7}
\multirow{2}{*}{$\rm{Q_{0}}$\,[GeV]} & \multirow{2}{*}{Data$\,\pm\,$(stat.)$\,\pm\,$(syst.)} & \multirow{2}{*}{MC@NLO} & P{\sc owheg} & P{\sc owheg} & \multirow{2}{*}{S{\sc herpa}} & A{\sc lpgen} & \multirow{2}{*}{$\rho^{\rm{i}}_{\rm{j}}$} \\
& & & +P{\sc ythia} & +H{\sc erwig} & & +H{\sc erwig} & \\
\toprule
\multicolumn{3}{l}{veto region: $|\rm{y}| < 0.8$} & & & & \\
25 & 76.9$\,\pm\,$1.1$\,^{+\,\mathrm{2.0}}_{-\,\mathrm{2.1}}$ & 79.5$\,\pm\,$0.1 & 75.0$\,\pm\,$0.3 & 74.3$\,\pm\,$0.3 & 74.9$\,\pm\,$0.3 & 76.7$\,\pm\,$0.3 & $\rho^{25}_{75}$ = 0.52\\
75 & 92.3$\,\pm\,$0.7$\,\pm\,$0.5 & 94.3 & 91.8$\,\pm\,$0.2 & 92.2$\,\pm\,$0.2 & 93.4$\,\pm\,$0.2 & 93.4$\,\pm\,$0.2 & $\rho^{75}_{150}$ = 0.51\\
150 & 97.8$\,^{+\,\mathrm{0.3}}_{-\,\mathrm{0.4}}$$\,\pm\,$0.4 & 98.4 & 97.2$\,\pm\,$0.1 & 97.6$\,\pm\,$0.1 & 97.8$\,\pm\,$0.1 & 98.0$\,\pm\,$0.1 & $\rho^{150}_{25}$ = 0.27\\
\toprule
\multicolumn{3}{l}{veto region: $0.8 \leq |\rm{y}| < 1.5$} & & & & \\
25 & 80.4$\,\pm\,$1.0$\,\pm\,$1.7 & 82.0$\,\pm\,$0.1 & 79.5$\,\pm\,$0.2 & 79.5$\,\pm\,$0.3 & 79.8$\,\pm\,$0.3 & 81.3$\,\pm\,$0.3 & $\rho^{25}_{75}$ = 0.49\\
75 & 93.9$\,\pm\,$0.6$\,^{+\,\mathrm{0.5}}_{-\,\mathrm{0.4}}$ & 94.7 & 93.5$\,\pm\,$0.2 & 93.8$\,\pm\,$0.2 & 94.8$\,\pm\,$0.1 & 94.7$\,\pm\,$0.2 & $\rho^{75}_{150}$ = 0.55\\
150 & 97.9$\,^{+\,\mathrm{0.3}}_{-\,\mathrm{0.4}}$$\,\pm\,$0.2 & 98.4 & 97.7$\,\pm\,$0.1 & 98.0$\,\pm\,$0.1 & 98.4$\,\pm\,$0.1 & 98.2$\,\pm\,$0.1 & $\rho^{150}_{25}$ = 0.29\\
\toprule
\multicolumn{3}{l}{veto region: $1.5 \leq |\rm{y}| < 2.1$} & & & & \\
25 & 86.8$\,^{+\,\mathrm{0.8}}_{-\,\mathrm{0.9}}$$\,^{+\,\mathrm{1.2}}_{-\,\mathrm{1.1}}$ & 86.1$\,\pm\,$0.1 & 85.4$\,\pm\,$0.2 & 85.5$\,\pm\,$0.2 & 85.6$\,\pm\,$0.2 & 86.4$\,\pm\,$0.2 & $\rho^{25}_{75}$ = 0.42\\
75 & 97.6$\,\pm\,$0.4$\,\pm\,$0.4 & 95.8 & 95.9$\,\pm\,$0.1 & 96.0$\,\pm\,$0.1 & 96.5$\,\pm\,$0.1 & 95.9$\,\pm\,$0.1 & $\rho^{75}_{150}$ = 0.48\\
150 & 99.4$\,^{+\,\mathrm{0.2}}_{-\,\mathrm{0.3}}$$\,\pm\,$0.2 & 98.8 & 98.7$\,\pm\,$0.1 & 98.8$\,\pm\,$0.1 & 98.9$\,\pm\,$0.1 & 98.8$\,\pm\,$0.1 & $\rho^{150}_{25}$ = 0.20\\
\toprule
\multicolumn{3}{l}{veto region: $|\rm{y}| < 2.1$} & & & & \\
25 & 56.4$\,\pm\,$1.3$\,^{+\,\mathrm{2.6}}_{-\,\mathrm{2.8}}$ & 57.0$\,\pm\,$0.1 & 52.7$\,\pm\,$0.3 & 52.5$\,\pm\,$0.3 & 54.0$\,\pm\,$0.3 & 55.2$\,\pm\,$0.3 & $\rho^{25}_{75}$ = 0.48\\
75 & 84.7$\,\pm\,$0.9$\,\pm\,$1.0 & 85.7$\,\pm\,$0.1 & 82.7$\,\pm\,$0.2 & 83.6$\,\pm\,$0.2 & 86.0$\,\pm\,$0.2 & 85.1$\,\pm\,$0.2 & $\rho^{75}_{150}$ = 0.50\\
150 & 95.2$\,^{+\,\mathrm{0.5}}_{-\,\mathrm{0.6}}$$\,\pm\,$0.4 & 95.6 & 93.9$\,\pm\,$0.1 & 94.5$\,\pm\,$0.1 & 95.3$\,\pm\,$0.1 & 95.1$\,\pm\,$0.1 & $\rho^{150}_{25}$ = 0.24\\
\toprule
\end{tabular}
\end{center}
\end{table*}
}

 {
\renewcommand{\arraystretch}{1.27}
\begin{table*}
\caption{
The measured values of $f(\Qsum)$~for $\Qsum=55, 150$~and $300$~GeV for the different rapidity intervals used to veto jet activity are presented, excluding any measurements of $f(\Qsum)=1.0$. The predictions from the Monte Carlo event generators and the statistical correlations ($\rho^{\rm i}_{\rm j}$) between measurements are presented in the same way as in Table \ref{t:sum:q0}. 
}\label{t:sum:qsum}

\begin{center}
\begin{tabular}{lccccccc}
\midrule
\multicolumn{8}{c}{$f(\Qsum)$ (\%)} \\
\cmidrule(l){2-7}
\multirow{2}{*}{$\rm{Q_{sum}}$\,[GeV]} & \multirow{2}{*}{Data$\,\pm\,$(stat.)$\,\pm\,$(syst.)} & \multirow{2}{*}{MC@NLO} & P{\sc owheg} & P{\sc owheg} & \multirow{2}{*}{S{\sc herpa}} & A{\sc lpgen} & \multirow{2}{*}{$\rho^{\rm{i}}_{\rm{j}}$} \\
& & & +P{\sc ythia} & +H{\sc erwig} & & +H{\sc erwig} & \\
\toprule
\multicolumn{3}{l}{veto region: $|\rm{y}| < 0.8$} & & & & \\
55 & 88.1$\,^{+\,\mathrm{0.8}}_{-\,\mathrm{0.9}}$$\,^{+\,\mathrm{1.3}}_{-\,\mathrm{1.4}}$ & 91.4$\,\pm\,$0.1 & 88.0$\,\pm\,$0.2 & 88.4$\,\pm\,$0.2 & 89.9$\,\pm\,$0.2 & 90.1$\,\pm\,$0.2 & $\rho^{55}_{150}$ = 0.45\\
150 & 97.4$\,^{+\,\mathrm{0.4}}_{-\,\mathrm{0.5}}$$\,^{+\,\mathrm{0.8}}_{-\,\mathrm{0.9}}$ & 98.4 & 97.2$\,\pm\,$0.1 & 97.6$\,\pm\,$0.1 & 97.8$\,\pm\,$0.1 & 98.0$\,\pm\,$0.1 & $\rho^{150}_{300}$ = 0.46\\
300 & 99.4$\,^{+\,\mathrm{0.2}}_{-\,\mathrm{0.3}}$$\,\pm\,$0.3 & 99.7 & 99.4 & 99.6 & 99.6 & 99.6 & $\rho^{300}_{55}$ = 0.20\\
\toprule
\multicolumn{3}{l}{veto region: $0.8 \leq |\rm{y}| < 1.5$} & & & & \\
55 & 89.3$\,\pm\,$0.8$\,\pm\,$0.9 & 92.0 & 90.6$\,\pm\,$0.2 & 91.1$\,\pm\,$0.2 & 92.2$\,\pm\,$0.2 & 92.0$\,\pm\,$0.2 & $\rho^{55}_{150}$ = 0.48\\
150 & 97.3$\,\pm\,$0.4$\,\pm\,$0.3 & 98.4 & 97.7$\,\pm\,$0.1 & 98.0$\,\pm\,$0.1 & 98.4$\,\pm\,$0.1 & 98.2$\,\pm\,$0.1 & $\rho^{150}_{300}$ = 0.34\\
300 & 99.6$\,^{+\,\mathrm{0.1}}_{-\,\mathrm{0.2}}$$\,\pm\,$0.1 & 99.8 & 99.6 & 99.6 & 99.7 & 99.6 & $\rho^{300}_{55}$ = 0.15\\
\toprule
\multicolumn{3}{l}{veto region: $1.5 \leq |\rm{y}| < 2.1$} & & & & \\
55 & 95.2$\,^{+\,\mathrm{0.5}}_{-\,\mathrm{0.6}}$$\,\pm\,$0.6 & 93.8 & 93.6$\,\pm\,$0.2 & 93.9$\,\pm\,$0.2 & 94.6$\,\pm\,$0.2 & 94.1$\,\pm\,$0.2 & $\rho^{55}_{150}$ = 0.40\\
150 & 99.3$\,^{+\,\mathrm{0.2}}_{-\,\mathrm{0.3}}$$\,\pm\,$0.2 & 98.8 & 98.7$\,\pm\,$0.1 & 98.8$\,\pm\,$0.1 & 98.9$\,\pm\,$0.1 & 98.8$\,\pm\,$0.1 & $-$ \\
\toprule
\multicolumn{3}{l}{veto region: $|\rm{y}| < 2.1$} & & & & \\
55 & 72.7$\,\pm\,$1.1$\,^{+\,\mathrm{2.3}}_{-\,\mathrm{2.5}}$ & 79.0$\,\pm\,$0.1 & 75.3$\,\pm\,$0.3 & 76.5$\,\pm\,$0.3 & 79.6$\,\pm\,$0.3 & 78.6$\,\pm\,$0.3 & $\rho^{55}_{150}$ = 0.47\\
150 & 92.1$\,\pm\,$0.7$\,\pm\,$0.8 & 95.6 & 93.9$\,\pm\,$0.1 & 94.5$\,\pm\,$0.1 & 95.3$\,\pm\,$0.1 & 95.1$\,\pm\,$0.1 & $\rho^{150}_{300}$ = 0.46\\
300 & 98.1$\,^{+\,\mathrm{0.3}}_{-\,\mathrm{0.4}}$$\,^{+\,\mathrm{0.2}}_{-\,\mathrm{0.3}}$ & 99.4 & 98.8$\,\pm\,$0.1 & 99.1$\,\pm\,$0.1 & 99.2$\,\pm\,$0.1 & 99.1$\,\pm\,$0.1 & $\rho^{300}_{55}$ = 0.21\\
\toprule
\end{tabular}
\end{center}
\end{table*}
}

 The precision of the data, coupled with the large spread of theory predictions, implies that higher-order theory predictions may be needed to describe the data in all regions of phase space. For example, the NLO plus parton shower predictions provided by \mcatnlo{} and \powheg{} have LO accuracy in the first parton emission and leading logarithmic (LL) accuracy for subsequent emissions. Similarly, the ME plus parton shower predictions provided by \sherpa{} and \alpgen{} are accurate to LO for the first three emissions and LL thereafter. Possible improvements on this accuracy include NLO calculations that account for the decay products of the top quarks \cite{Denner:2010jp,Bevilacqua:2010qb} and calculations of $t\bar{t}+j(j)$ at NLO \cite{Dittmaier:2007wz,Dittmaier:2008uj,Melnikov:2010iu,Bevilacqua:2010ve,Bevilacqua:2011aa,Melnikov:2011qx}.
 
\section{Conclusions}
\label{sec-conclusions}

Precision measurements of the jet activity in $\ttbar$ events were performed using proton-proton collisions recorded by the ATLAS detector at the LHC. The $\ttbar$ events were selected in the dilepton decay channel with two identified $b$-jets. Events were subsequently vetoed if they contained an additional jet with transverse momentum above a threshold, $Q_0$, in a central rapidity interval. The fraction of $t\bar{t}$ events that survive the jet veto was presented as a function of $Q_0$ for four different central rapidity interval definitions. An alternate measurement was also performed, in which the $t\bar{t}$ events were vetoed if the scalar transverse momentum sum of the additional jets in each rapidity interval was above a defined threshold, $\Qsum$.

The data were fully corrected for detector effects and compared to the predictions from state-of-the-art MC event generators. \mcatnlo{}, \powheg{}, \alpgen{} and \sherpa{} are observed to give a reasonable description of the data, when the additional jets are vetoed in the rapidity interval $|y|<2.1$. However, all four generators predict too much jet activity in the most forward rapidity interval, $1.5 \leq |y| < 2.1$. Furthermore, \mcatnlo{} produces too little activity in the central region $|y|<0.8$.

The data were compared to the predictions obtained after increasing (or decreasing) the amount of initial state radiation produced by the \pythia{} parton shower when applied to \acermc{} events. These initial state parton shower variations have been used to determine modelling uncertainties in previous ATLAS top quark measurements. Although the data are within the band of these predictions, the size of the band is a factor of two or more larger than the experimental precision. The results presented here can be used to constrain model-dependent uncertainties in future measurements.

\section{Acknowledgements}

We thank CERN for the very successful operation of the LHC, as well as the
support staff from our institutions without whom ATLAS could not be
operated efficiently.

We acknowledge the support of ANPCyT, Argentina; YerPhI, Armenia; ARC,
Australia; BMWF, Austria; ANAS, Azerbaijan; SSTC, Belarus; CNPq and FAPESP,
Brazil; NSERC, NRC and CFI, Canada; CERN; CONICYT, Chile; CAS, MOST and NSFC,
China; COLCIENCIAS, Colombia; MSMT CR, MPO CR and VSC CR, Czech Republic;
DNRF, DNSRC and Lundbeck Foundation, Denmark; EPLANET and ERC, European Union;
IN2P3-CNRS, CEA-DSM/IRFU, France; GNAS, Georgia; BMBF, DFG, HGF, MPG and AvH
Foundation, Germany; GSRT, Greece; ISF, MINERVA, GIF, DIP and Benoziyo Center,
Israel; INFN, Italy; MEXT and JSPS, Japan; CNRST, Morocco; FOM and NWO,
Netherlands; RCN, Norway; MNiSW, Poland; GRICES and FCT, Portugal; MERYS
(MECTS), Romania; MES of Russia and ROSATOM, Russian Federation; JINR; MSTD,
Serbia; MSSR, Slovakia; ARRS and MVZT, Slovenia; DST/NRF, South Africa;
MICINN, Spain; SRC and Wallenberg Foundation, Sweden; SER, SNSF and Cantons of
Bern and Geneva, Switzerland; NSC, Taiwan; TAEK, Turkey; STFC, the Royal
Society and Leverhulme Trust, United Kingdom; DOE and NSF, United States of
America.

The crucial computing support from all WLCG partners is acknowledged
gratefully, in particular from CERN and the ATLAS Tier-1 facilities at
TRIUMF (Canada), NDGF (Denmark, Norway, Sweden), CC-IN2P3 (France),
KIT/GridKA (Germany), INFN-CNAF (Italy), NL-T1 (Netherlands), PIC (Spain),
ASGC (Taiwan), RAL (UK) and BNL (USA) and in the Tier-2 facilities
worldwide.

\bibliographystyle{spphys}       % APS-like style for physics

\bibliography{topjetveto}

\clearpage
\onecolumn
% ATLAS Collaboration author list for 13-FEB-2012
% Data extracted on 14-Mar-2012 for paperid 199
%\documentclass[11pt]{article}
%\usepackage{a4wide}\begin{document}
\begin{flushleft}
{\Large The ATLAS Collaboration}

\bigskip

G.~Aad$^{\rm 48}$,
B.~Abbott$^{\rm 112}$,
J.~Abdallah$^{\rm 11}$,
S.~Abdel~Khalek$^{\rm 116}$,
A.A.~Abdelalim$^{\rm 49}$,
A.~Abdesselam$^{\rm 119}$,
O.~Abdinov$^{\rm 10}$,
B.~Abi$^{\rm 113}$,
M.~Abolins$^{\rm 89}$,
O.S.~AbouZeid$^{\rm 159}$
% ...
\bigskip

% ...
$^{10}$ Institute of Physics, Azerbaijan Academy of Sciences, Baku, Azerbaijan\\
$^{11}$ Institut de F\'isica d'Altes Energies and Departament de F\'isica de la Universitat Aut\`onoma  de Barcelona and ICREA, Barcelona, Spain\\
% ...
$^{48}$ Fakult\"{a}t f\"{u}r Mathematik und Physik, Albert-Ludwigs-Universit\"{a}t, Freiburg i.Br., Germany\\
$^{49}$ Section de Physique, Universit\'e de Gen\`eve, Geneva, Switzerland\\
% ...
$^{89}$ Department of Physics and Astronomy, Michigan State University, East Lansing MI, United States of America\\
% ...
$^{112}$ Homer L. Dodge Department of Physics and Astronomy, University of Oklahoma, Norman OK, United States of America\\
$^{113}$ Department of Physics, Oklahoma State University, Stillwater OK, United States of America\\
% ...
$^{116}$ LAL, Univ. Paris-Sud and CNRS/IN2P3, Orsay, France\\
% ...
$^{119}$ Department of Physics, Oxford University, Oxford, United Kingdom\\
% ...
$^{159}$ Department of Physics, University of Toronto, Toronto ON, Canada\\
% ...
\end{flushleft}

%\end{document}

\end{document}